\documentclass[12pt,preprint]{aastex}
\usepackage{color}
\usepackage{ulem}

\usepackage{graphicx,longtable}
\usepackage{array}
\usepackage{amsmath}
\usepackage{amssymb}
\usepackage{subfigure}

\def\Msun{~M_{\odot} }

\begin{document}
\title{Revisiting Impacts of Nuclear Burning for Reviving Weak Shocks in 
Neutrino-Driven Supernovae}

\author{Ko Nakamura\altaffilmark{1,2}, 
Tomoya Takiwaki\altaffilmark{3}, 
Kei Kotake\altaffilmark{4,1}, and Nobuya Nishimura\altaffilmark{5,6}}

\altaffiltext{1}{Division of Theoretical Astronomy, National Astronomical Observatory of Japan,
2-21-1 Osawa, Mitaka, Tokyo, 181-8588, Japan}
\altaffiltext{2}{Faculty of Science and Engineering, Waseda University, Phkubo 3-4-1, Shinjuku, Tokyo, 169-8555, Japan}
\altaffiltext{3}{Center for Computational Astrophysics, National Astronomical Observatory of Japan,
2-21-1 Osawa, Mitaka, Tokyo, 181-8588, Japan}
\altaffiltext{4}{Department of Applied Science, Fukuoka University, 
8-19-1, Nanakuma, Jonan-ku, Fukuoka, 814-0180, Japan}
\altaffiltext{5}{Astrophysics Group, iEPSAM, Keele University, Keele, ST5 5BG, UK}
\altaffiltext{6}{Department of Physics, University of Basel, Klingelbergstrasse 82, 4056 Basel, Switzerland}

\begin{abstract}
We revisit potential impacts of nuclear burning on the onset
 of the neutrino-driven explosions of core-collapse supernovae.
By changing the neutrino luminosity and its decay time
to obtain parametric explosions
in one-(1D) and two-dimensional (2D)
models with or without a 13-isotope $\alpha$ network, 
we study how
the inclusion of nuclear burning could affect
 the postbounce dynamics for four progenitor models; three 
for $15.0 \Msun$ stars, one for an $11.2 \Msun$ star.
We find that the energy supply due to nuclear burning of 
infalling material behind the shock can energize the
shock expansion especially for models that produce only 
marginal explosions in the absence of nuclear burning. 
These models are energized by nuclear energy deposition
when the shock front passes through the silicon-rich layer 
and/or later it touches the oxygen-rich layer.
Depending on the neutrino luminosity and its decay time, 
a diagnostic energy of explosion increases 
up to a few times $10^{50}$ erg 
 for models with nuclear burning compared to the corresponding models 
without.
We point out that these features are most remarkable for the Limongi-Chieffi progenitor in both 1D and 2D, because 
the progenitor model possesses a massive oxygen layer with its inner-edge radius being smallest among the employed progenitors, so that the shock can touch 
the rich fuel on a shorter timescale after bounce.
The energy difference is generally smaller ($\sim 0.1-0.2 \times 10^{51}$ erg) in 2D than in 1D (at most $\sim 0.6 \times 10^{51}$ erg). 
This is because neutrino-driven convection and the shock instability
 in 2D models enhance the neutrino heating efficiency, which makes
the contribution of nuclear burning relatively smaller compared to
1D models.
Considering uncertainties in progenitor models, our results indicate
that nuclear burning should remain as one of the important
ingredients to foster the onset of neutrino-driven explosions.
\end{abstract}
\medskip


\maketitle

\vskip 1.3cm


\section{Introduction}
Ever since the dawn of modern core-collapse supernova (CCSN) theory, 
the neutrino-heating mechanism \citep{colgate},
 in which a supernova shock is revived by neutrino energy deposition
to trigger explosions \citep{wils85,bethe85},
has been the leading candidate for the explosion  mechanism
  for more than four decades.
 However, the simplest, spherically-symmetric (1D) form of this 
mechanism fails,  except for super-AGB stars 
at the low-mass end \citep{bernhard12a}, to explode canonical massive stars 
\citep{Rampp00,lieb01,thom03,Sumiyoshi05}.
 Pushed by accumulating supernova observations of the blast 
morphology 
 \citep[e.g.,][and references therein]{wang08,tanaka12},
a number of multi-dimensional (multi-D)
hydrodynamic simulations have been reported so far,
which gives us a confidence that hydrodynamic motions associated with convection 
\citep[e.g.,][]{Herant92,Burrows95,Janka96,frye02,fryer04a} 
and the Standing-Accretion-Shock-Instability 
\citep[SASI, e.g., ][]{Blondin03,Scheck04,scheck06,Ohnishi06,ohnishi07,ott_multi,Murphy08,Foglizzo06,thierry07,endeve12,thierry12,iwakami1,iwakami2,rodrigo09_2,rodrigo09,rodrigo10,hanke,rodrigo13} 
can help the onset of neutrino-driven explosions
\citep[see collective references in][]{thomas12,kotake12}.

In fact, neutrino-driven explosions have been obtained
in first-principle two-(2D) and three-(3D) dimensional simulations 
in which spectral neutrino transport is solved by various 
approximations \citep[e.g.,][]{kotake12b}.
The Garching group 
\citep{Buras06a,Buras06b,Marek09, hanke13, bernhard11,bernhard12a,bernhard12b,bernhard13} 
included one of the best available neutrino transfer approximations 
by the ray-by-ray variable Eddington factor method.
The Oak Ridge group \citep{bruenn09,bruenn13} 
included a ray-by-ray multi-group flux-limited diffusion transport 
with the best available weak interactions.
The Nippon group\footnote{"Nippon" stands for Japan in Japanese, and from 
 now on we like to call our team as so whose members come all around Japan.} \citep{Suwa10,Suwa11,suwa12,Takiwaki11,takiwaki13} 
employed a ray-by-ray isotropic diffusion source approximation \citep{idsa} 
with a reduced set of weak interactions\footnote{See \citet{sumiyoshi12} for collective references about detailed neutrino transport schemes.}.

This success, however, is accompanying further new question. 
The explosion energies obtained in 
some 2D models are underpowered by up to a factor of 10
compared to the canonical supernova kinetic energy 
($\sim 10^{51}$ erg, see table 1 in \citet{Kotake11} for a current 
summary).  What on earth is missing furthermore ? 
3D hydrodynamics has been 
pointed out to boost the onset of neutrino-driven explosions compared to 2D
\citep{nordhaus10}, although it
 is still under considerable debate \citep{hanke,hanke13,Takiwaki11,couch12}.
Very recently, general relativity has been reported
to help the onset of multi-D neutrino-driven explosions by
 \citet{bernhard11,bernhard12b} in 2D simulations with detailed neutrino transport and by 
\citet{kuroda12,kuroda13} in 3D simulations but with approximate neutrino transport.
Impacts of nuclear equations of state (EOSes) have been
investigated in multi-D simulations by \citet{Marek09,marek09b,suwa12} and  \citet{couch13}. 
However, there 
may still remain further room to study more detailed nuclear physical
impacts in these first principle multi-D simulations, such as the 
density dependence of symmetry energy and the skewness of compressibility 
\citep{steiner,lattimer12} and influences of light nuclei 
\citep[e.g.,][]{sumi08,arcones,nakamura}
and of inelastic neutrino-nucleus scattering 
\citep[e.g.,][]{haxton,ohnishi07,langanke08}
on enhancing the neutrino heating rates in the gain region. 
More recently, impacts of improved neutrino interactions based on
 the 1D full Boltzmann simulations have been elaborately investigated
 \citep{lentz11, lentz12}.
The neutrino-driven mechanism would be assisted by 
other candidate mechanisms such as the acoustic mechanism 
\citep[e.g.,][]{burr06}
or the magnetohydrodynamic mechanism (e.g., 
\citet{kota04a,kota04b,taki04,taki09,burr07,fogli_B,martin11,taki_kota}, 
see also \citet{kota06} for collective references therein). 
Other possibilities include QCD phase transitions
in the core of the protoneutron star
\citep[e.g.,][]{takahara88,sage09}
viscous heating by the 
magnetorotational instability \citep{thomp05},
or energy
dissipation via Alfv\'en waves \citep{suzu08}.

Joining in these efforts to look for some possible ingredients to 
foster explosions,
we pay attention to the roles of nuclear burning in this study.
To the best of our knowledge, \citet{janka01a} were 
the first to clearly point out 
that an additional energy released by nuclear burning of 
infalling material behind the shock could make a significant 
contribution to affect the explosion  energy (see their Eq.(5)). 
The mass in the silicon (Si) layer, depending sensitively on the 
progenitor masses and structures, is in the range of 
$\sim 0.3 - 0.6 M_{\sun}$ \citep{WW95,woos02,limongi}.
Since the release of nuclear energy 
in Si burning is $\approx 10^{18}~{\rm erg \, g}^{-1}$, 
a few $10^{50}$ erg are expected to be 
deposited by the explosive nuclear burning. 
It should be noted that nuclear burning 
has been included in the full-scale simulations by the Garching group
 \citep{rampp02,Buras06a,
Buras06b,Marek09}, in which composition changes of silicon, oxygen 
(and similarly neon and magnesium), and carbon
and their nuclear energy release are computed by a ``flashing" treatment
 \citep[see Appendix \ref{app-flash}, and also Appendix B.2 in ][]{rampp02}. In a series of 
 multi-D simulations in which neutrino transport is treated by a 
 more approximative way to follow a long-term postbounce 
evolution in the context 
 of the neutrino-driven mechanism, nuclear burning
 is included by a small network calculation \citep[e.g.,][]{kifo,scheck06,annop,hammer,arcones_2011,ugliano}. 
However in
these literatures, impacts of nuclear 
burning on the supernova dynamics have not been unambiguously 
investigated so far.
 In conference proceedings, the Oak Ridge group reported 
2D explosion models based on their radiation-hydrodynamic simulations 
\citep{bruenn06,mezza07} for 11.2$\Msun$ and 15.0$\Msun$ stars, 
only when an alpha network calculation was included, but not when 
they applied the flashing treatment.
They pointed out that oxygen burning 
 assists the (weak) shock
to move farther out due to the additional pressure support 
in the vicinity of the weak shock. These situations motivate 
us to revisit the impacts of nuclear burning on assisting
the shock propagation by performing hydrodynamic simulations
including a network calculation.

 In the present work, we take the following strategy to clearly
see the roles of nuclear burning. Firstly 
we try, in the spirit of 
\citet{burogoshy} and \citet{Janka01},
to find a critical condition in 1D, in which nuclear burning affects 
the criteria of explosion. Instead of performing full-scale 
radiation-hydrodynamic simulations which are computationally expensive, 
we employ a light-bulb scheme to trigger explosions 
\citep[e.g.,][]{Janka96} for the sake of our systematic 
survey.
Previously the role of nuclear burning seems to be considered as 
negligible using a very limited set of progenitor models
but we will show that for a previously untested progenitor model, nuclear burning
can really push the weak shock farther out to help explosions.

This paper opens with the description of numerical setup including 
 information about our hydrodynamic code with nuclear network and 
about initial models (Section 2). Results are given in section 3. 
After giving a detailed explanation in 1D models 
how nuclear burning could affect the postbounce dynamics 
(section \ref{sec-1d}, \ref{sec-prog}), 
we move on to discuss our 2D models to study how nuclear burning 
interacts with multi-D hydrodynamics (section \ref{sec-2d}). 
We summarize our results and discuss their implications in Section 4.

\section{Numerical Setup}

\subsection{Hydrodynamics with Nuclear Network}
We solve the hydrodynamic equations corresponding to the conservation of mass, 
momentum, and energy,
\begin{equation}\label{eq-mass}
\frac{d \rho}{d t} + \rho \nabla \cdot \mathbf{v} = 0,
\end{equation}
\begin{equation}\label{eq-momentum}
\rho \frac{d \mathbf{v}}{d t} = - \nabla p - \rho \nabla \Phi,
\end{equation}
\begin{equation}\label{eq-energy}
\frac{\partial e}{\partial t} + \nabla \cdot \left[ (e+p)\mathbf{v} \right] = - p \mathbf{v} \cdot \nabla \Phi + \rho (H - C + Q),
\end{equation}
where $\rho$ is the mass density, $\mathbf{v}$ the fluid velocity, $p$ the pressure, 
$\Phi$ the gravitational potential, and $e$ the total energy density, respectively.
The Lagrangian derivative is denoted by $d/dt \equiv \partial / \partial t + \mbox{\boldmath$v$} \cdot \nabla$. To treat Newtonian self-gravity, a monopole approximation is employed. The tabulated realistic equation of state based on the relativistic mean field theory \citep{Shen98} is
implemented according to the prescription in \citet{kotake}.
The term $Q$ in Eq. (\ref{eq-energy}) denotes the net energy deposition rate by nuclear burning.
The goal of this paper is to explore the effect of this term on shock revival. We compare two cases:
(a) For {\it burning} case, we estimate $Q$ by calculating a simple nuclear reaction network, and 
(b) for {\it non-burning} case,
we do not solve the nuclear network ($Q=0$) but adopt Shen EOS throughout simulations. 

For {\it burning} case we are keeping track of 13 species of $\alpha$ network
(from $^4$He to $^{56}$Ni) by solving a separate advection equation for each species.
The nuclear reaction network is mainly based on the REACLIB database \citep{rauscher00}.
Experimentally determined masses \citep{audi} and reactions \citep{angulo} are adopted 
if available. It should be noted that our network does not include the 
photodissociation of iron elements because Shen EOS adopted in this study 
takes account of these endothermic effects.
Note also that we solve the reaction network only for the grids 
where $T<5 \times 10^9$ K, assuming that above this temperature 
the local chemical composition is in nuclear statistical 
equilibrium (NSE).

In this study, we employ the so-called light-bulb scheme \citep{Janka96},
in which neutrino heating and cooling is adjusted parametrically to trigger explosions.
Following \citet{Janka01} and \citet{nordhaus10}, 
the neutrino heating ($H$) and cooling rates ($C$) 
in Eq. (\ref{eq-energy})
are given by,
\begin{equation}\label{eq-heating}
\begin{split}
H = 1.544 \times 10^{20} 
& \left( \frac{L_{\nu_{\rm e}}}{10^{52} \, {\rm erg \, s^{-1}}}\right) \left( \frac{T_{\nu_{\rm e}}}{4 \, {\rm MeV}} \right)^2 \\
& \times  \left( \frac{r}{100 \, {\rm km}} \right)^{-2} (Y_{\rm n} + Y_{\rm p}) \, e^{-\tau_{\nu_{\rm e}}}~[{\rm erg}~{\rm g}^{-1}~{\rm s}^{-1}],
\end{split}
\end{equation}
\begin{equation}\label{eq-cooling}
C = 1.399 \times 10^{20} \left( \frac{T}{2 \, {\rm MeV}} \right)^6 (Y_{\rm n} + Y_{\rm p}) \, e^{-\tau_{\nu_{\rm e}}}, [{\rm erg}~{\rm g}^{-1}~{\rm s}^{-1}],
\end{equation}
where $L_{\nu_{\rm e}}$ is the electron-neutrino luminosity that is 
assumed to be equal to the anti-electron neutrino luminosity 
($L_{\bar{\nu}_{\rm e}} = L_{\nu_{\rm e}}$), $T_{\nu_{\rm e}}$ is 
the electron neutrino temperature assumed to be kept constant as $4$ MeV, $r$ 
is the distance from the center, $T$ is the local fluid temperature, $Y_n$ and $Y_p$
are the neutron and proton fractions, and $\tau_{\nu_e}$ is the electron neutrino
optical depth that we estimate according to Eq. (7) in \citet{hanke}. 

Note in this study that neutrino luminosity is assumed to evolve 
exponentially with time \citep{kifo}
as 
\begin{equation}
L_{\nu_{\rm e}} = L_{\bar{\nu}_{\rm e}} = L_{\nu0} \, {\rm exp}
(- t_{\rm pb}
/t_{d}),
\label{nuL}
\end{equation}
where $L_{\nu0}$ denotes the initial luminosity, 
$t_{\rm pb}$ is the time measured after core bounce
$t_{d}$ is the decay time, respectively.
$L_{\nu0}$ and $t_{d}$ are treated as free parameters. Note that 
neutrino heating 
and cooling are switched on only after core bounce.

Only after core bounce, neutrino heating and cooling is switched on,
according to the prescriptions (Equations (\ref{eq-heating}) and 
(\ref{eq-cooling})) 
assuming $L_{\nu_{\rm e}} = L_{\bar{\nu}_{\rm e}}$ and $T_{\nu_{\rm e}} = T_{\bar{\nu}_{\rm e}}$. Before bounce, we employ the $Y_e$ 
prescription proposed by \citet{simple}, in which $Y_e$ is given simply 
as a function of density, and after that, we 
refrain from solving the change of $Y_e$ following 
\citet{Murphy08,nordhaus10} and \citet{hanke} 
(see, however, \citet{Ohnishi06}). As for the hydro-solver,
we employ the ZEUS-MP code \citep{hayes} which has been modified for
core-collapse simulations \citep[e.g.,][]{iwakami1,iwakami2}.
The computational grid is comprised of 300 logarithmically spaced, radial zones 
from the center up to 5000 km. 
For 2D models we adopt coarse mesh points ($n_{\theta} = 32$ uniform grids) in the polar direction to make it possible to perform 174 models
 in 2D covering a wide range of the parameter region. For some selected 
models, a finer resolution ($n_{\theta} = 128$) is taken.

In order to induce non-spherical instability after the stall of the prompt 
bounce shock,
we have added a radial velocity perturbation, $\delta v_r(r, \theta, \phi)$, to 
the steady spherically symmetric flow according to the following equation,
\begin{equation}
v_r(r, \theta, \phi) = v_r^{0}(r) +\delta v_r(\theta, \phi),
\end{equation}
with
\begin{equation}
\delta v_r = 0.01 \times {\rm rnum} \times \, v_r^0 (r,\theta) \, ,
\end{equation}
where $v_r^0 (r,\theta)$ is the unperturbed radial velocity and $\delta v_r$ is the random multi-mode perturbation with a random number $-1 < {\rm rnum} < 1$.

\begin{figure}[htpb]
\begin{center}
\includegraphics[scale=0.6]{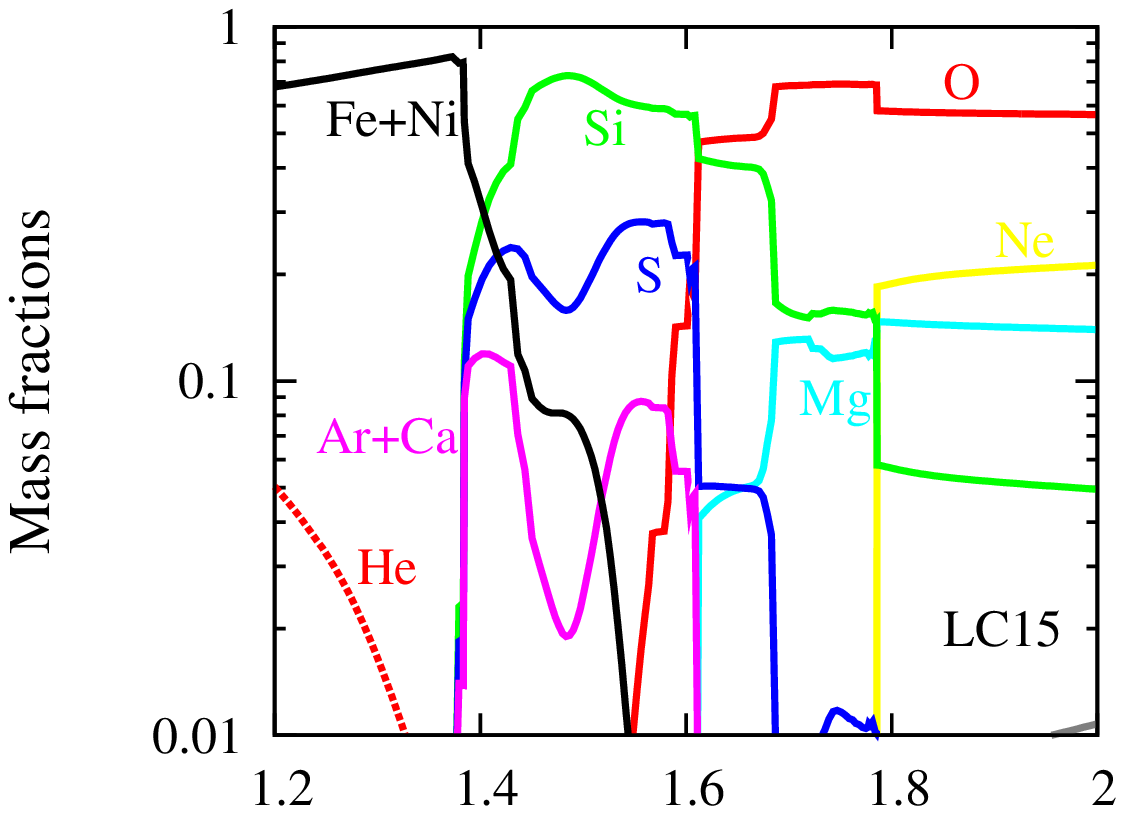}
\includegraphics[scale=0.6]{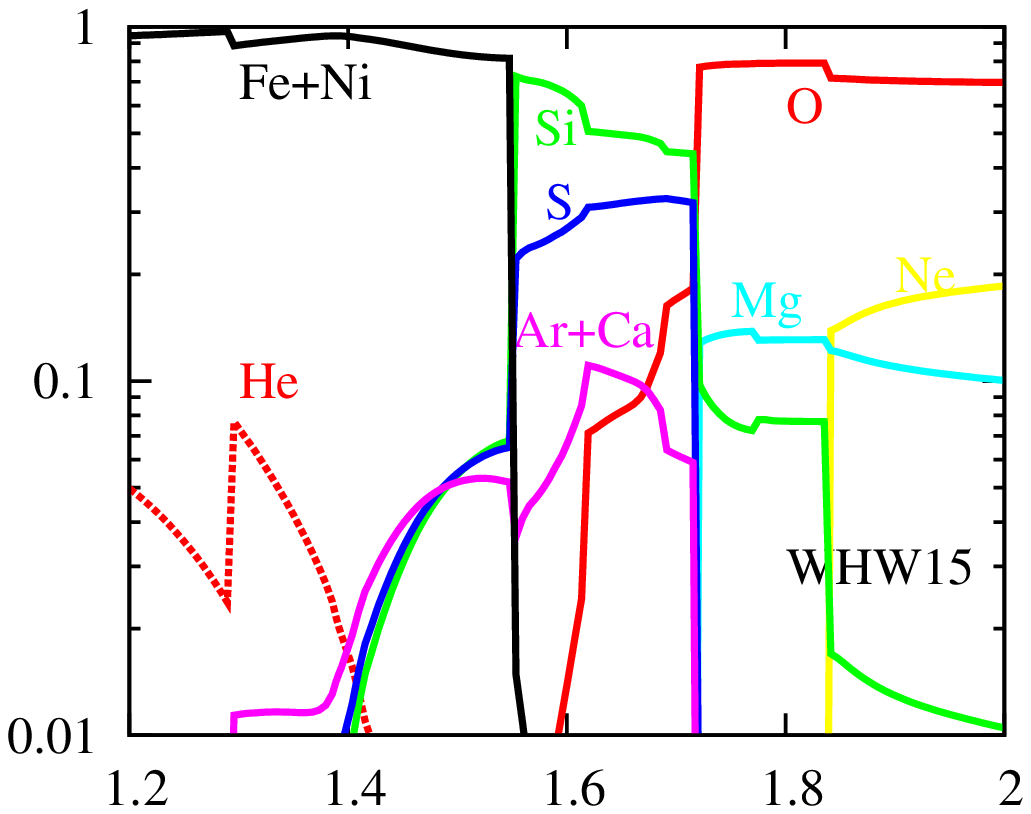}
\includegraphics[scale=0.6]{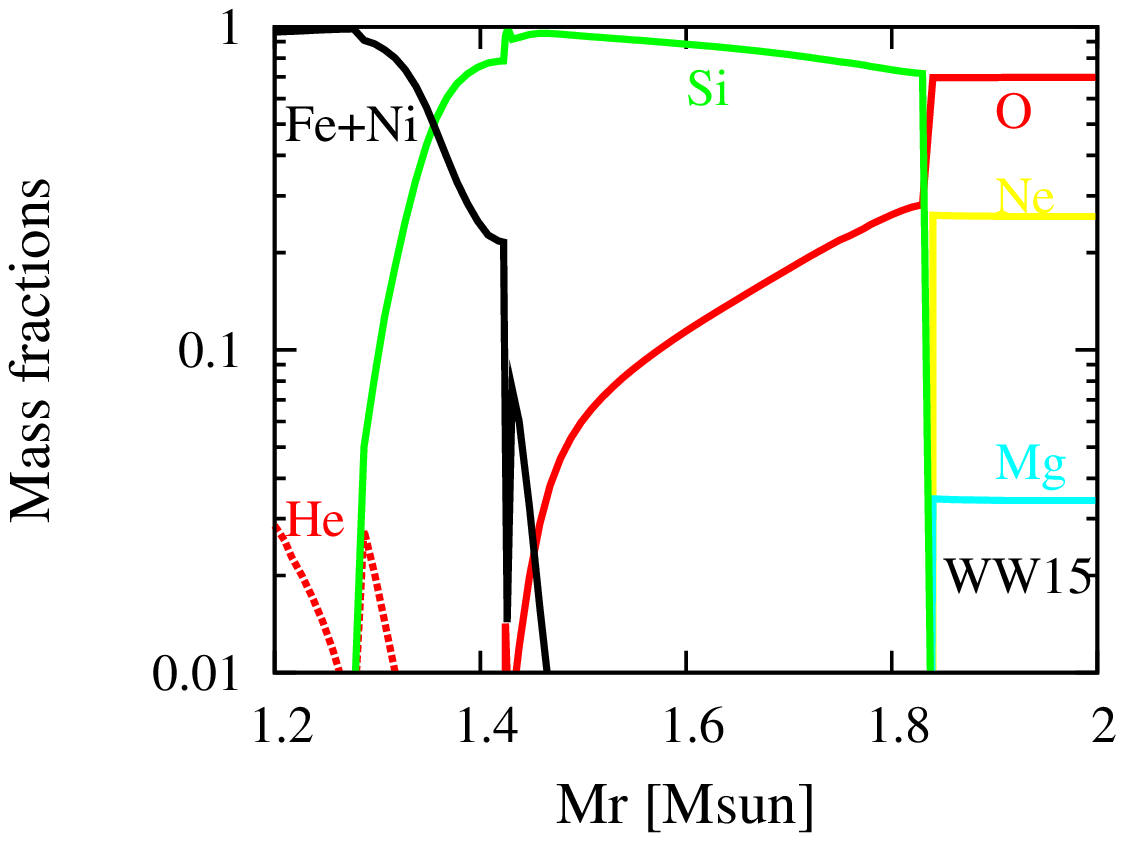}
\includegraphics[scale=0.6]{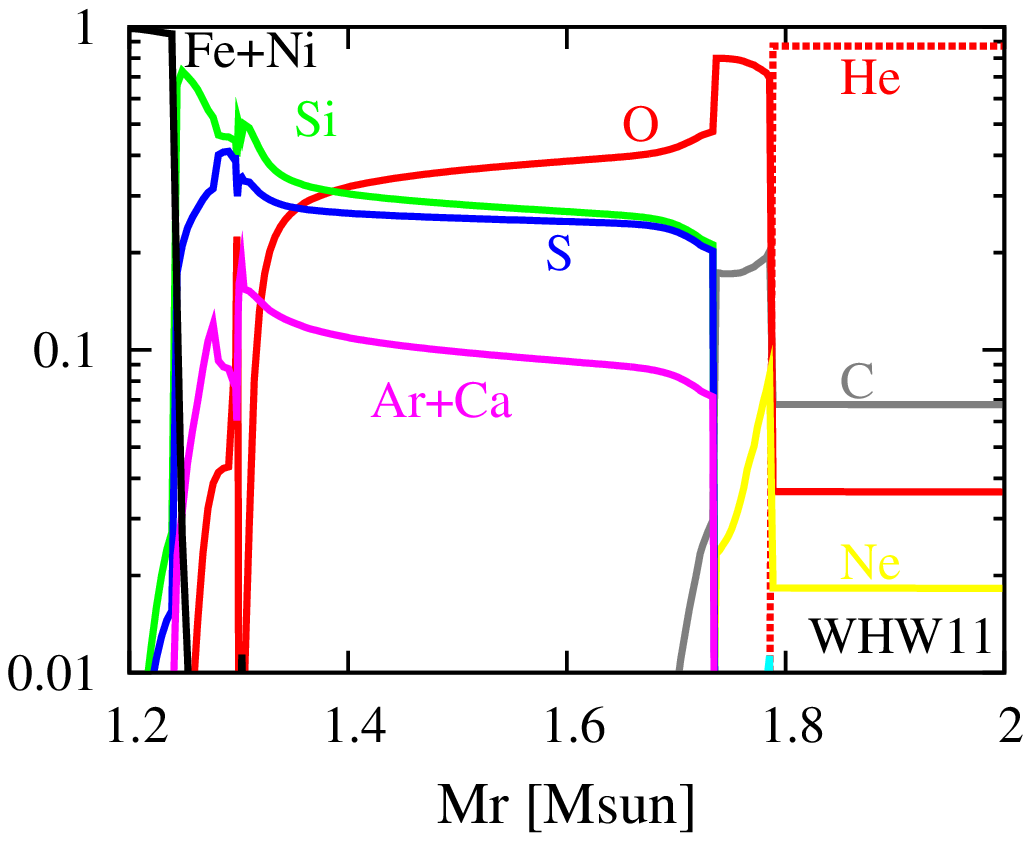}
\end{center}
\caption{Precollapse composition distributions for the 15 $M_{\sun}$ stars 
of \citet{limongi} (labeled by LC15, {\it top left}), 
\citet{WW95} (WW15, {\it bottom left}), and 
\citet{woos02} (WHW15, {\it top right}) and for the $11.2 \Msun$ star 
of \citet{woos02} (WHW11, {\it bottom right}).}
\label{fig-pmodel}
\end{figure}

\subsection{Progenitor models}
In this study, we employ four progenitor models;
three for $15.0 \Msun$ stars of
\citet[][hereafter LC15]{limongi}, \citet[][WW15]{WW95}, 
and  \citet[][WHW15]{woos02} and one for an $11.2 \Msun$ star of 
\citet[][WHW11]{woos02}. 
For all the models, Figure \ref{fig-pmodel}
shows the precollapse composition profiles near from the outer edge of 
the iron core to outside. As we will 
explain in the next section, burning of the oxygen shell behind 
the (weakly propagating) shock plays an important 
role in assisting the shock expansion.
Therefore, the earlier the oxygen layer touches the (stalling)
shock after bounce, the better it could work. 
Among the three variants of the
15 $M_{\odot}$ progenitors, the inner edge of the oxygen layer 
(seen as a sharp decline in solid red lines of Figure \ref{fig-pmodel})
is positioned much closer to the center for models LC15 (closest, top
left panel) and WHW15 (next closest, 
top right panel) compared to model WW15 (bottom left panel).
Table 1 shows a summary of the precollapse abundance distributions,
 in which each quantity from the left to right column corresponds to
 the different progenitor models, the progenitor
mass, the mass of the iron core,
 the outer edge of the iron core, the mass of the
  silicon layer, the outer edge of the silicon layer, and the mass of
the oxygen layer, respectively.  
The edge between each layer is defined as the radius where the most abundant element shifts to one another (see, Figure \ref{fig-pmodel}).
The mass of oxygen layer for the 15
$M_{\odot}$ models of LC15 and WHW15 ($M_{O}$ in the table) is larger
than the other progenitors (i.e., WW15 and WHW11) and their oxygen layers 
(denoted by $R_{{\rm Si}/{\rm O}}$) are positioned much 
closer to the center, so that they can touch the supernova shock
in a shorter timescale after bounce 
(before the neutrino luminosity gets smaller with time).
As one would anticipate, the impacts of nuclear burning are most 
remarkable for the LC15 progenitor as we will show in the later sections.

\begin{table}[htbp]
\caption{Summary of progenitor models and their composition features
(see
 text for the definition of each quantity).}
\begin{tabular}{ccccccl}
\hline
Model & $M_{\rm total}$   & $M_{\rm Fe}$ & $R_{\rm Fe/Si}$ & $M_{\rm Si}$ & $R_{\rm Si/O}$ & $M_{\rm O}$\\
          & $(\Msun)$           & $(\Msun)$       & $(10^3 \,{\rm km})$ & $(\Msun)$& $(10^3 \,{\rm km})$ & $(\Msun)$\\
\hline
LC15     & 13.4                   & 1.44              & 1.31                    & 0.221         & 2.22               & 0.814 \\
WHW15 & 12.6                  & 1.55               & 1.96                    & 0.124            & 2.97                & 0.943   \\
WW15 & 15.0                    & 1.42                & 1.31                   & 0.436           & 7.44                & 0.649 \\
WHW11& 10.8                  & 1.24                & 1.00                   & 0.168             & 3.74                & 0.289    \\
\hline
\\
\\
\end{tabular}
\label{tbl-model}
\end{table}

\clearpage
\section{Results}

\begin{figure}[htpb]
\begin{center}
\includegraphics[scale=0.4]{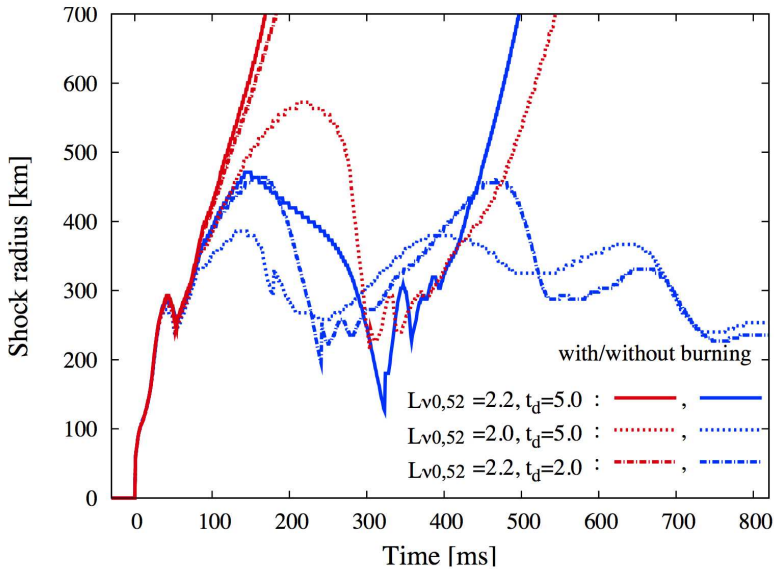}
\end{center}
\caption{Time evolution of the shock radii for model LC15
with different initial neutrino luminosities ($L_{\nu0, 52}$ in unit of
 $10^{52}$ erg s$^{-1}$) and the decay time
 ($t_d$ in unit of s). The red and blue line corresponds to the
 results with 
  and
  without the energy feedback from nuclear
reactions, respectively.}
\label{fig-rsh1d}
\end{figure}

In section 3.1, we start to investigate how the energy feedback from
nuclear burning could affect the postbounce dynamics in 1D simulations.
 Then we study how the nuclear-burning impacts are sensitive to the progenitor models,
namely by the precollapse structures and their composition profiles
   (section 3.2). In section 3.3, we then move on to discuss how
nuclear burning would affect the 2D dynamics.

\subsection{Impact of Nuclear Burning in 1D simulations}\label{sec-1d}

Relying on the light-bulb scheme in this study, the destiny of the
stalling bounce shock (whether it will revive or not) depends simply
on the two parameters; the input neutrino
luminosity $L_{\nu0}$ and the decay time $t_{d}$ 
(see Equation (\ref{nuL}))\footnote{Without seeing a shock revival in $\sim$ 1 s
postbounce, we call it as ''non-exploding'' in this study}. Note in the
following that we characterize models as ($L_{\nu0,52}, t_{d}) = (x, y)$ for
convenience, in which the luminosity and the decay time is $x \times
10^{52}$ (erg/s) and $y$ (s), respectively.   

Figure \ref{fig-rsh1d} shows comparisons of the postbounce shock evolution in 1D LC15 models 
depending on the two parameters ($L_{\nu0,52}, t_{d}$) 
and nuclear energy feedback from $\alpha$ network calculation. 
Chosen three sets of parameters: ($L_{\nu0,52}, t_{d}$) = (2.2, 5.0), (2.0, 5.0), and (2.2, 2.0),
are shown as a solid line, a dotted line, and a dash-dotted line, respectively. 
All of the models with nuclear burning, marked with red lines in Figure \ref{fig-rsh1d}, 
present a shock expansion leading to explosions, while among three models without nuclear burning 
(blue lines) only the model with relatively higher luminosity and longer decay time 
(($L_{\nu0,52}$, $t_d$) = (2.2, 5.0), solid blue line) exhibits a shock revival.
Note that in all the six cases 
in Figure \ref{fig-rsh1d}, the bounce shock 
stalls and then transits to a passive shock 
which presents negative radial velocity behind the shock.
And only after that, the additional energy gain due to
nuclear burning acts to bifurcate the path of the passive shock, namely
whether the shock experiences recession afterward
(for all the blue lines in Figure \ref{fig-rsh1d}) or expansion (for red lines)
with different revival timescales depending on the input neutrino 
parameters.

As seen from Figure \ref{fig-rsh1d}, larger input neutrino luminosity and shorter
decay timescale unsurprisingly leads to easier explosions.
More importantly, by comparing dotted red with dotted blue line 
(($L_{\nu0,52}$, $t_d$) 
= (2.0, 5.0)), the shock is shown to shift from recession to 
expansion by the inclusion of nuclear burning. 
In the case of higher neutrino luminosity ($L_{\nu0,52}$ = 2.2), 
the trajectories of the shock are observed to be rather similar when the effect of nuclear burning is taken into account
(compare solid red with dashed red line). 

\begin{figure}[htbp]
\begin{center}
   \begin{tabular}{cc}
\resizebox{75mm}{!}{\includegraphics{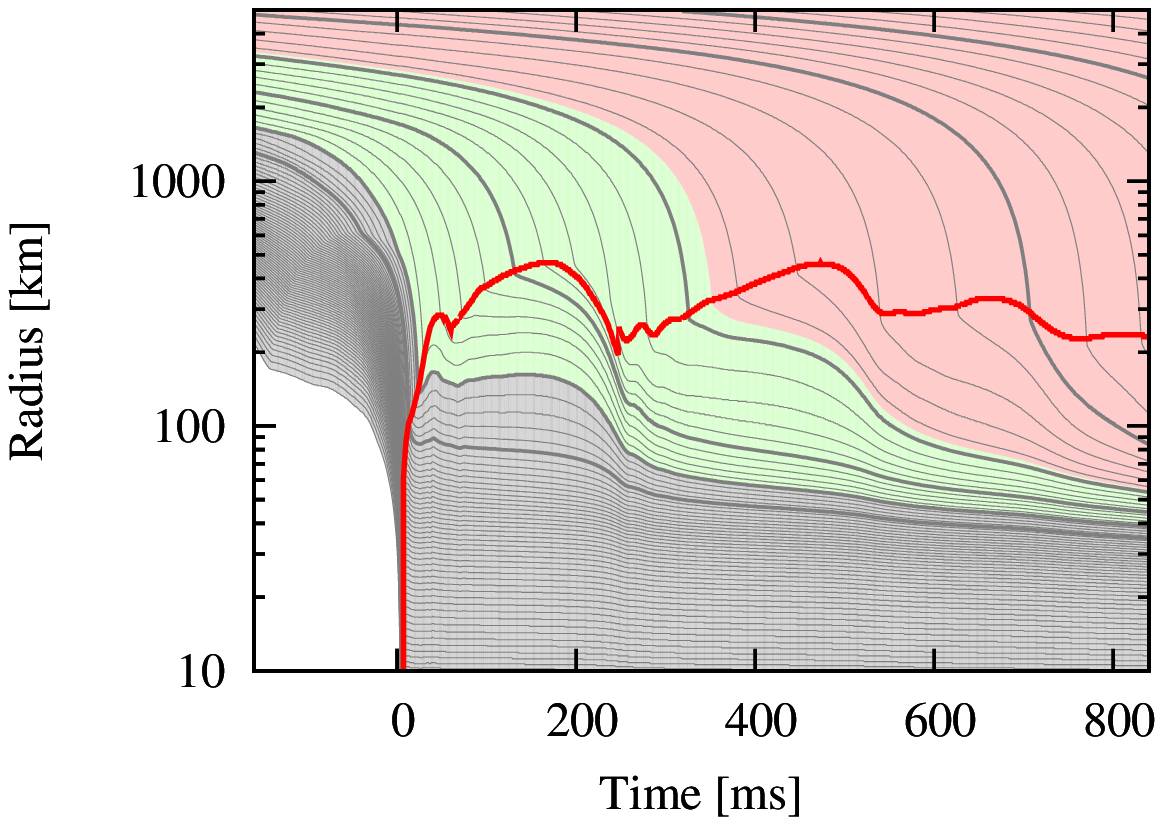}} &
\resizebox{75mm}{!}{\includegraphics{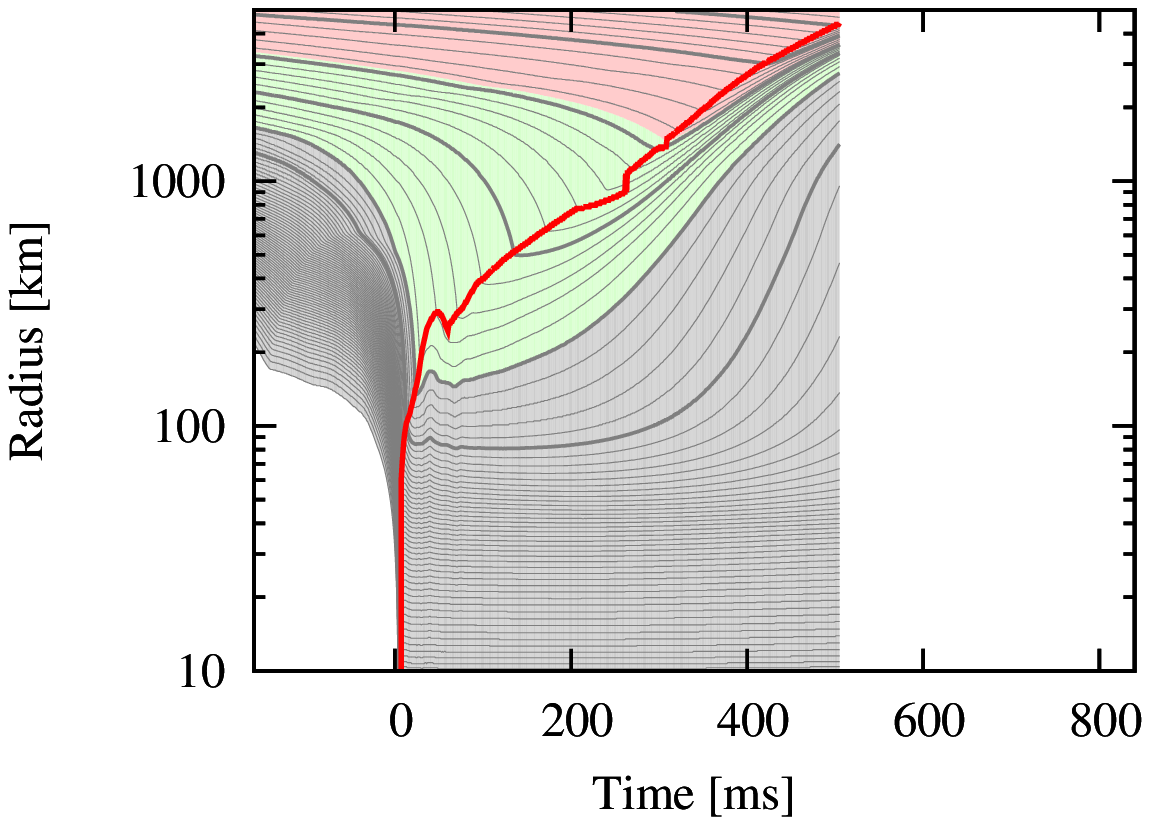}}  \\
   \end{tabular}
\end{center}
\caption{Evolution of model LC15
with a parameter set of $(L_{\nu 0,52}, t_{d}) = (2.2, 2.0)$ visualized by the mass-shell
trajectories. The thick red line starting at $t=0$ denotes the position
of the shock.  Both cases either without ({\it left}) or with
({\it right}) the energy feedback from nuclear reactions are shown.
The regions colored by gray, green, and red correspond to
 the iron, silicon, and oxygen layers, respectively.
Thick gray lines correspond to the mass coordinates from 1.3 to 1.8$\Msun$ with every 0.1 $M_{\odot}$
(thin gray lines with every $0.02 M_{\odot}$).}
\label{mshell-m15-2220}
\end{figure}

\begin{figure}[htbp]
\begin{center}
   \begin{tabular}{cc}
\resizebox{75mm}{!}{\includegraphics{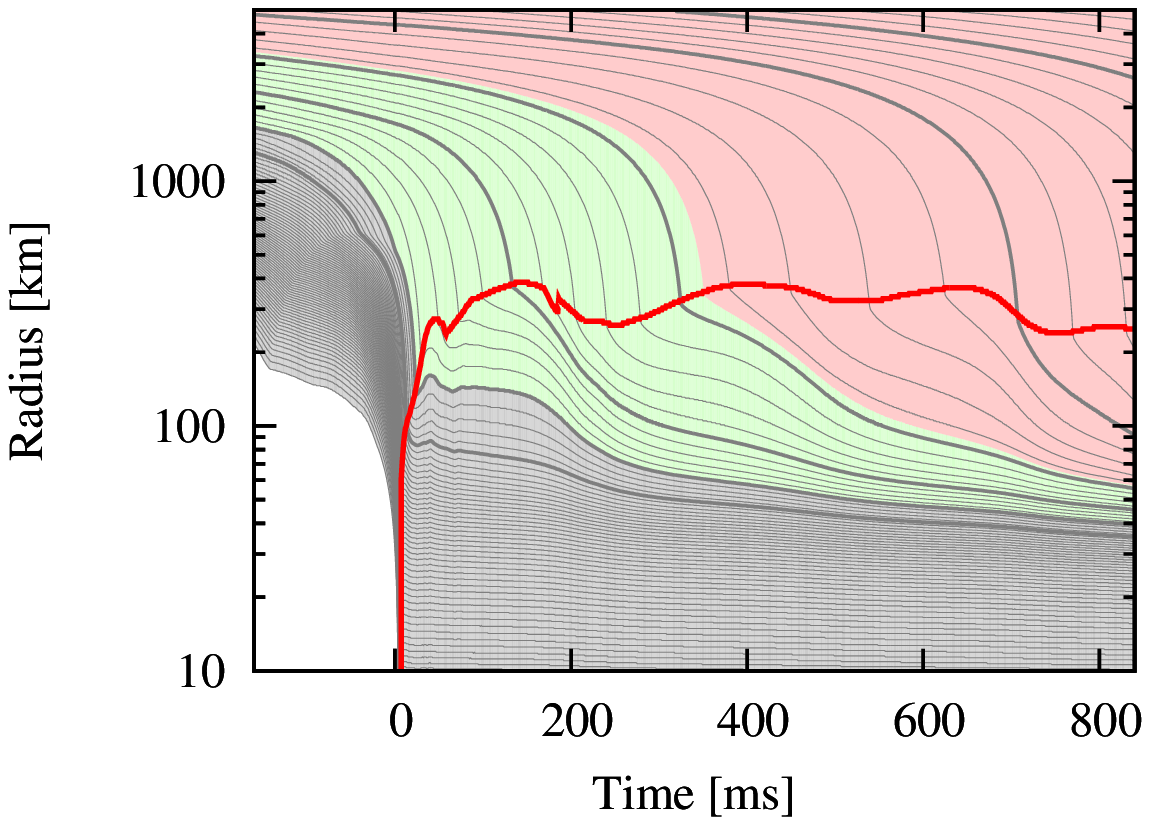}} &
\resizebox{75mm}{!}{\includegraphics{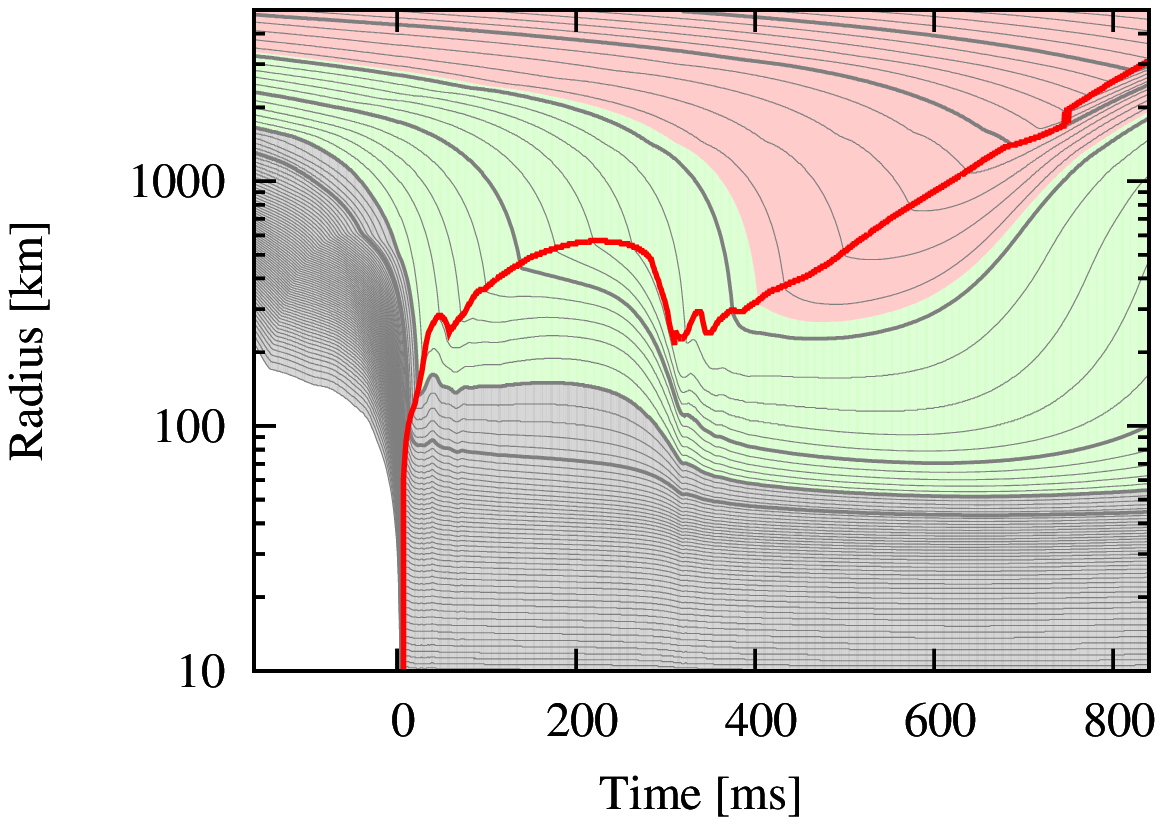}}  \\
   \end{tabular}
\end{center}
\caption{Same as Figure \ref{mshell-m15-2220} but for
the parameter set of $(L_{\nu 0,52}, t_d) = (2.0, 5.0)$.}
\label{mshell-m15-2050}
\end{figure}

In the following, we elaborate on how and why the
shock expansion is affected by nuclear burning as observed in Figure \ref{fig-rsh1d}. 
Figure \ref{mshell-m15-2220} and \ref{mshell-m15-2050} show
the
mass-shell trajectory of models LC15 with the different parameter set of
$(L_{\nu 0,52}, t_{d}) = (2.2, 2.0)$ and $(L_{\nu 0, 52}, t_{d}) = (2.0, 5.0)$, respectively.
Note here that the former and latter case corresponds to the dashed and
dotted line in Figure \ref{fig-rsh1d}. Without nuclear burning 
(left panels in Figure \ref{mshell-m15-2220} and \ref{mshell-m15-2050}), 
the stalled shock oscillates, but 
does not turn into expansion (see also Figure \ref{fig-rsh1d}).
With nuclear burning, 
right panels of Figures \ref{mshell-m15-2220} and \ref{mshell-m15-2050}, 
the shock expansion can be seen to take place 
when 
the shock front passes through
the Si-rich layer (see the behavior of the thick red line in the
green region in
the right panel of Figure \ref{mshell-m15-2220})
or later it touches the O-rich layer (e,g., the shock in the red
region in the right panel of Figure
\ref{mshell-m15-2050}). For the latter case, the bounce shock firstly stalls as
in the non-burning model (compare the left with the right panel in
Figure \ref{mshell-m15-2050}), but then the shock front 
deviates from
the non-burning case when it encounters with the
O-rich layer.

\begin{figure}[htbp]
\begin{center}
\includegraphics[scale=0.7]{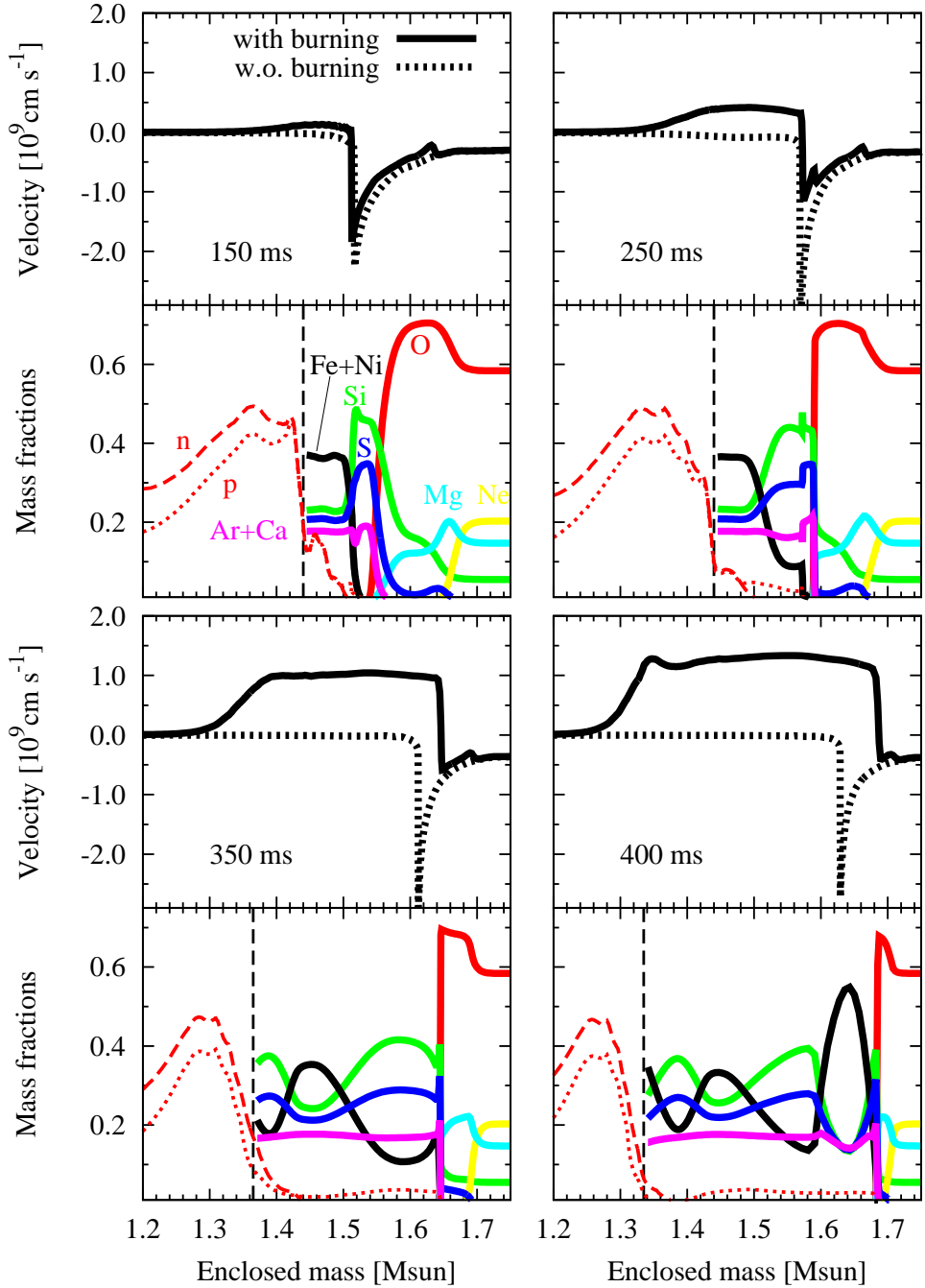}
\end{center}
\caption{Snapshots of velocity profile ({\it top}) and composition
distribution ({\it bottom}) for model LC15 with 
$(L_{\nu 0,52}, t_d) = (2.2, 2.0)$
at selected postbounce epochs ($t_{\rm pb} =
150, 250, 350,$ and $400$ ms). Solid and dotted line 
in the top of each panel shows the velocity profile with or without
nuclear burning, respectively. In the bottom part, distributions of
representative elements of the burning model are shown. 
Note that the abundances of neutron
(n) and proton (p) are estimated from Shen EOS and the others are 
calculated from the nuclear network calculation.
Nucleon-rich region in the abundance plot is separated by a vertical dashed line.
}
\label{snap1dm15_2220}
\end{figure}

\begin{figure}[htbp]
\includegraphics{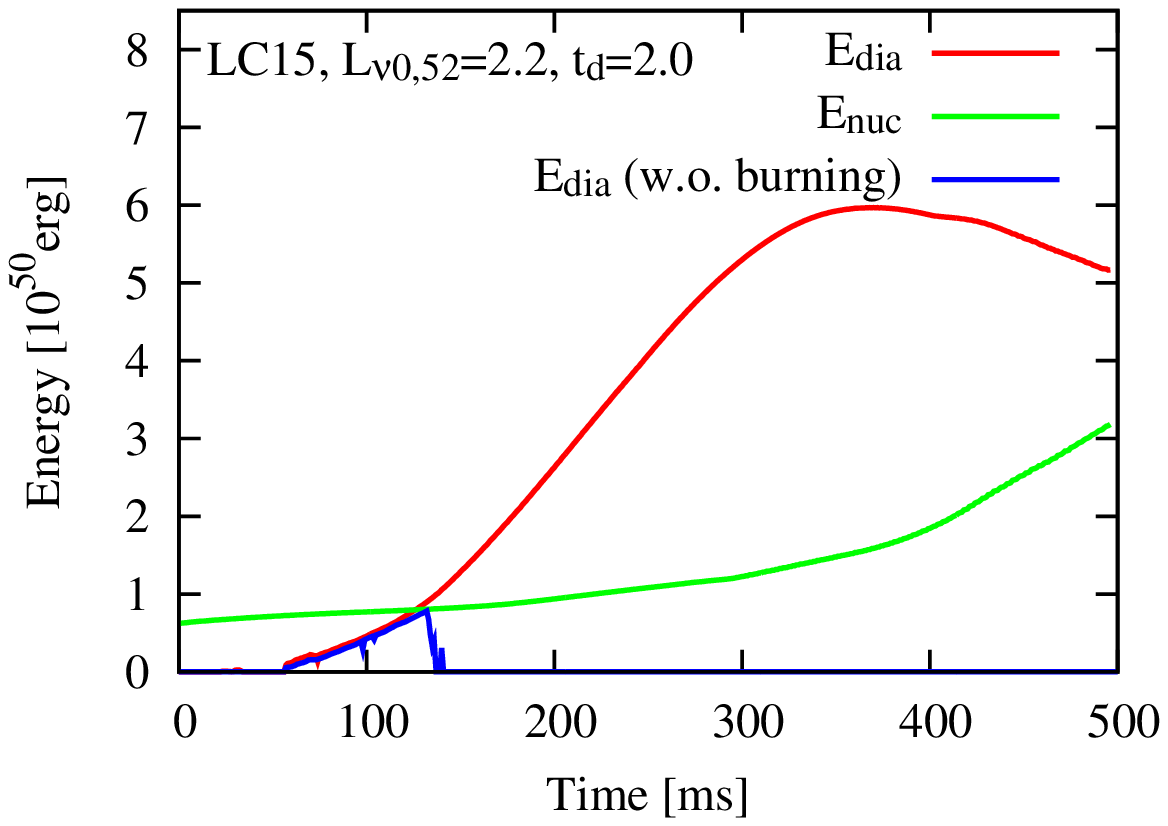}
\caption{
Time evolution of diagnostic energy is shown for the same model as Figure \ref{snap1dm15_2220}.
The green line is the energy released by 
nuclear burning and the diagnostic energy without nuclear
burning ({\it blue}) is also shown. Here the diagnostic energy is
defined as the sum of the kinetic,
thermal, and gravitational energy of fluid elements with positive radial velocity.}
\label{eex1d_m15_2220}
\end{figure}

\clearpage
To look more in detail how the nuclear burning contributes to the shock
acceleration, Figure \ref{snap1dm15_2220} shows the radial velocity profiles and the
composition distributions for model LC15 with the parameter set 
$(L_{\nu 0,52}, t_{d}) = (2.2, 2.0)$ (the same parameter set as in Figure \ref{mshell-m15-2220}).
At $t_{\rm pb} = 150$ ms (top left panel), the shock front
is in the progenitor silicon-rich layer. 
Behind the shock front, 
heavier elements are synthesized as shown.
The nuclear energy released by silicon burning 
heats the material behind the shock, 
making it have a small positive velocity there 
(compare the velocity profiles with and without nuclear burning in the top left panel).
The difference between the velocity profiles with versus without nuclear
burning becomes outstanding when
the oxygen-rich layer starts to touch the shock front ($t_{\rm pb} > 250$ ms).
Figure \ref{eex1d_m15_2220} shows the evolution of
the {\it diagnostic} (explosion) energy 
for burning (red line) and non-burning (blue line) cases
and also the net energy released via nuclear reactions (green
line). As in \citet{Suwa10}, we define a {\it diagnostic} energy 
that refers to the integral of the energy over all zones that 
have a positive sum of the specific internal, kinetic, and gravitational energy. 
It is impossible to calculate the final energy of the explosion that is still occurring at this early post-bounce stage.
After silicon burning starts to feed energy behind the
shock in addition to
neutrino heating (in the gain region, e.g., $t_{\rm pb}
= 150 $ ms, see the top left panel in Figure \ref{snap1dm15_2220}), 
the diagnostic energy deviates from the one without burning 
(compare red with blue line 
in Figure \ref{eex1d_m15_2220}), 
which is also clearly visible in the shock evolution 
(Figure \ref{mshell-m15-2220}).
From 
Figure \ref{eex1d_m15_2220}, the total amount of $3.1 \times 10^{50}$ erg 
is shown to be released through nuclear burning in this case, 
lifting up the diagnostic energy to be $5.0 \times 10^{50}$ erg.

\begin{figure}[htbp]
\begin{center}
\includegraphics[scale=0.8]{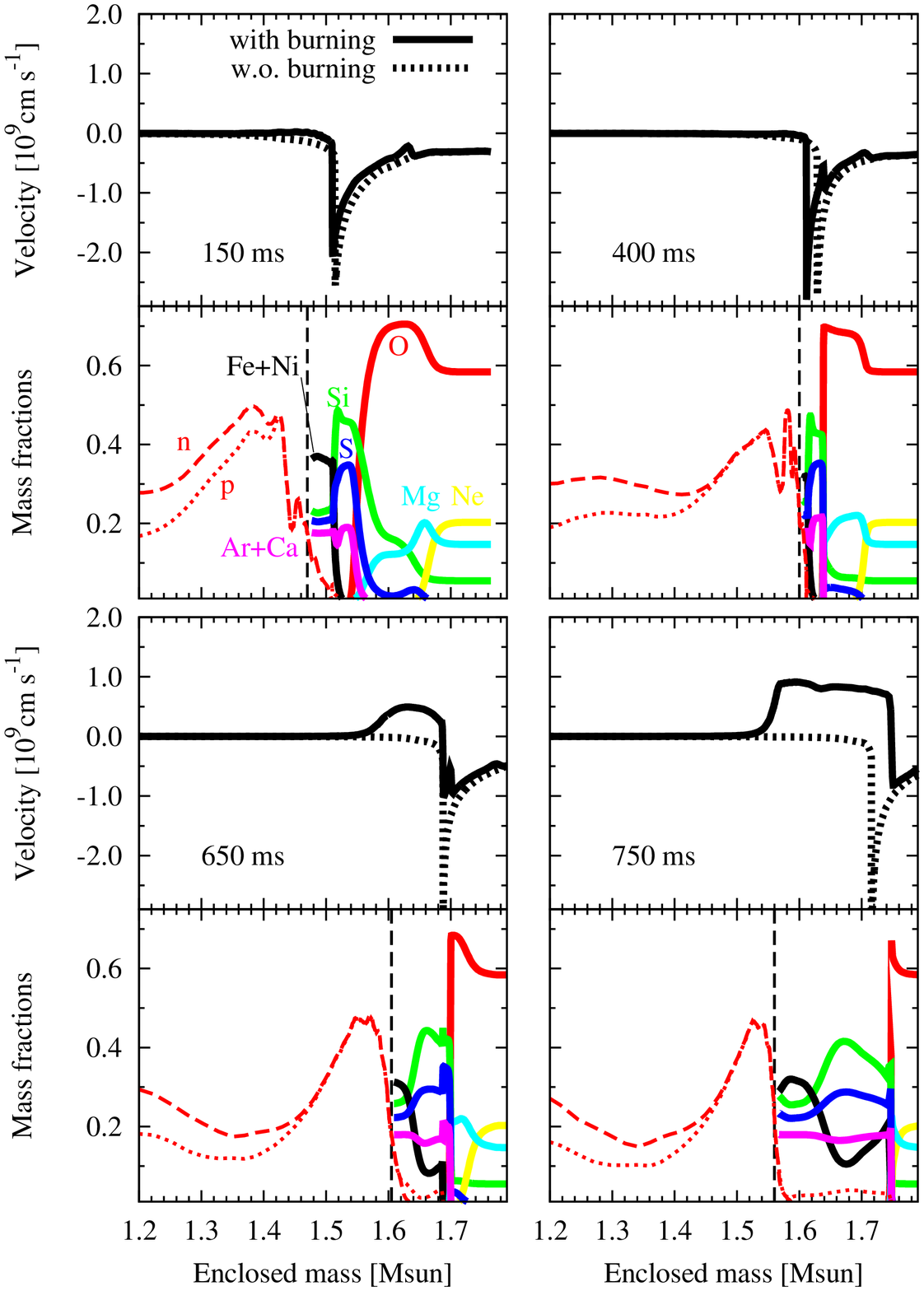}
\end{center}
\caption{Same as Figure \ref{snap1dm15_2220} but for the parameter set of 
$(L_{\nu 0,52}, t_d) = (2.0, 5.0)$.}
\label{snap1dm15_2050}
\end{figure}

\begin{figure}[htbp]
\includegraphics{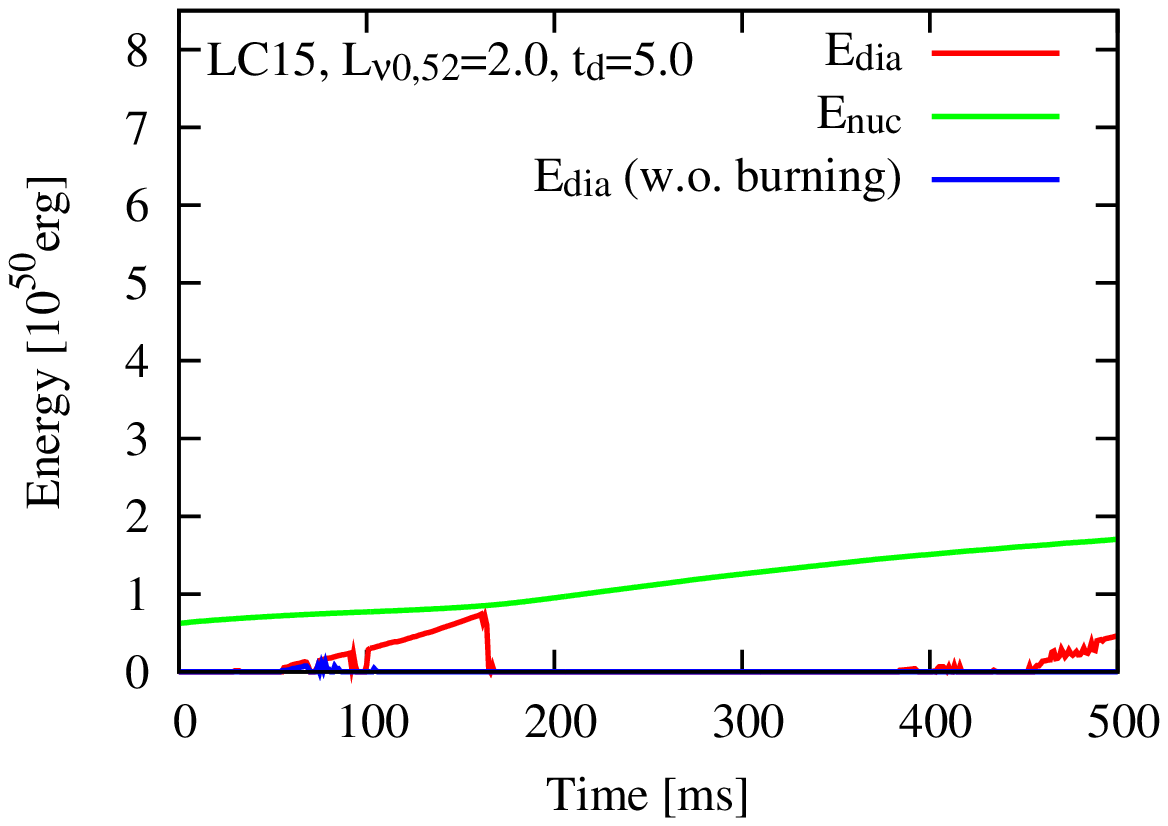}
\caption{Same as Figure \ref{eex1d_m15_2220} but for the parameter set of 
$(L_{\nu 0,52}, t_d) = (2.0, 5.0)$.}
\label{eex1d_m15_2050}
\end{figure}

\begin{figure}[htbp]
\includegraphics{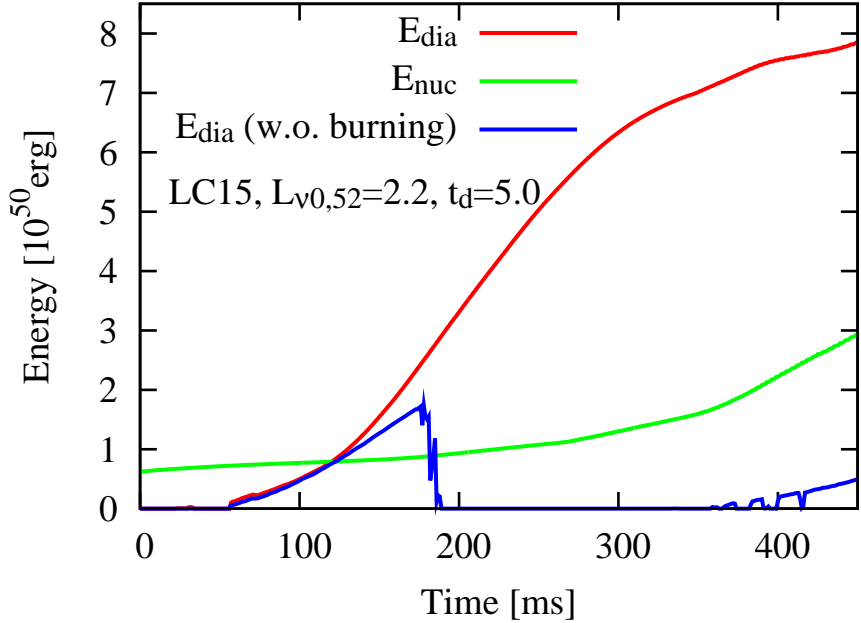}
\caption{Same as Figure \ref{eex1d_m15_2220} but 
for the most energetic case, ($L_{\nu 0,52}, t_d) = (2.2, 5.0$), 
among the three examples shown 
in Figure \ref{fig-rsh1d}. The diagnostic energy with nuclear burning is 
about $ 8.0 \times
10^{50}$ ergs at $t_{\rm pb} = 465$ ms and still keeps rising.
The net energy released via nuclear reactions at this time is
$\sim 3.0 \times 10^{50}$ erg, occupying a significant fraction ($\sim$
40\%) of the diagnostic energy. For the model without nuclear burning,
the diagnostic energy is $\sim 0.8 \times 10^{50}$ erg at that time
and closely saturates to be $E_{\rm dia} \sim 2.3 \times 10^{50}$ erg
afterward.}
\label{eex1d_m15_2250}
\end{figure}

As we already mentioned, oxygen burning predominantly 
triggers the shock
expansion for the parameter set taken for Figure \ref{mshell-m15-2050}. But also in this case, the silicon layer is shown to be burned as a heating source 
(top left panel of Figure \ref{snap1dm15_2050}), 
which is the reason that the shock position becomes larger 
compared to the non-burning model 
(Figure \ref{mshell-m15-2050}). 
When the shock front begins to swallow the oxygen layer 
at $\sim 400$ ms postbounce 
(the right panel of Figure \ref{mshell-m15-2050}), 
the fresh fuel supplies energy to assist the shock expansion 
(see, from top right, bottom left, to bottom right panels of Figure \ref{snap1dm15_2050}). 
If not for the energy gain, 
the stalled shock does not revive earlier than $t=750$ ms 
as seen from 
the left panel of Figure \ref{mshell-m15-2050}.
Even with the aid of nuclear burning, the explosion for this model 
(Figure \ref{eex1d_m15_2050}) 
is weaker ($\lesssim 10^{50}$ erg) compared to the more luminous models 
(Figures \ref{eex1d_m15_2220} and \ref{eex1d_m15_2250}). 
This suggests that nuclear burning has a secondary impact on the
explosion mechanism 
--- it can assist explosions only when neutrino heating is working enough
strong to push the weak shock to the fuel layers.

\clearpage

\subsection{Progenitor Dependence}\label{sec-prog}

\begin{figure}[htpb]
\begin{center}
\includegraphics[scale=0.8]{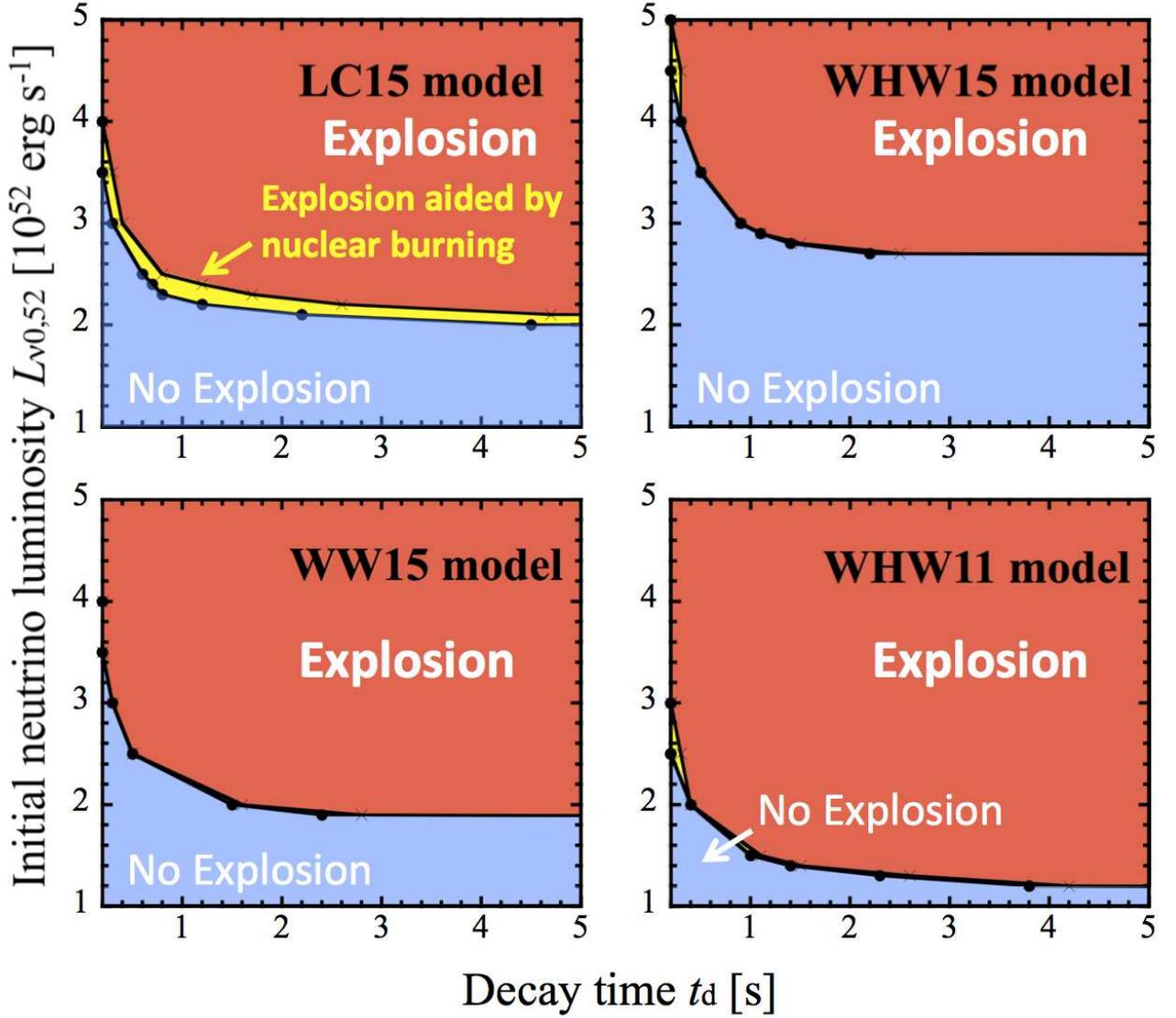}
\end{center}
\caption{Parameter maps of the initial neutrino luminosity $L_{\nu0,52}$
and its decay time $t_d$ that separates the non-exploding regime
(blue region) from the exploding one (red region) in 1D
simulations for the four different progenitors. A horizontal yellow
region in-between (clearly visible for the LC15 progenitor;{\it top left})
 shows the parameter region in which 1D explosions are obtained when
the network calculation is performed. }
\label{fig-para1d}
\end{figure}

Figure \ref{fig-para1d} shows a parameter map on the ($L_{\nu0,52}$, $t_d$) 
plane for each progenitor. As can be seen, higher neutrino luminosity 
and/or longer decay time leads to easier explosions
 (colored by red and denoted as "Explosion"), while is simply opposite 
for smaller neutrino luminosity and/or shorter decay time 
(colored by light-blue and denoted as "No explosion" in the figure). 
For the LC15 progenitor model (top left panel),
 a parameter region colored by yellow can be seen between the 
exploding and the non-exploding regime, in which 
 an explosion is obtained only when nuclear burning is included 
in the hydrodynamics simulations. The emergence of the yellow region 
means that the minimum neutrino luminosity necessary to drive an 
explosion is reduced by taking into account energy feedback form
nuclear burning. 
The burning-mediated regime is clearly visible only for the LC15
progenitor. As already mentioned in section 2.2, this is because
 this model possesses a massive oxygen layer
and the oxygen shell is
positioned closest to the center among the progenitors taken in this study.

The area of the yellow region in Figure \ref{fig-para1d} is not so large even for the LC15 progenitor, 
which suggests again that nuclear burning has a secondary importance.
In the case of energetic explosions, for example ($L_{\nu0,52}, t_d$) = (3.0, 0.8), 
a difference of diagnostic energy between models with and without nuclear burning is 
$\sim 5 \times 10^{49}$ erg (Table \ref{tbl-result}). 
In the case of marginal weak explosions with $E_{\rm dia} \lesssim 10^{50}$ erg (which is often the case in recent first-principle CCSN simulations), however,
it should be emphasized that the inclusion of nuclear burning could
 increase the diagnostic energy up to about $\sim 0.6 \times 10^{51}$ erg.

\begin{figure}[htpb]
\begin{center}
\includegraphics[scale=0.8]{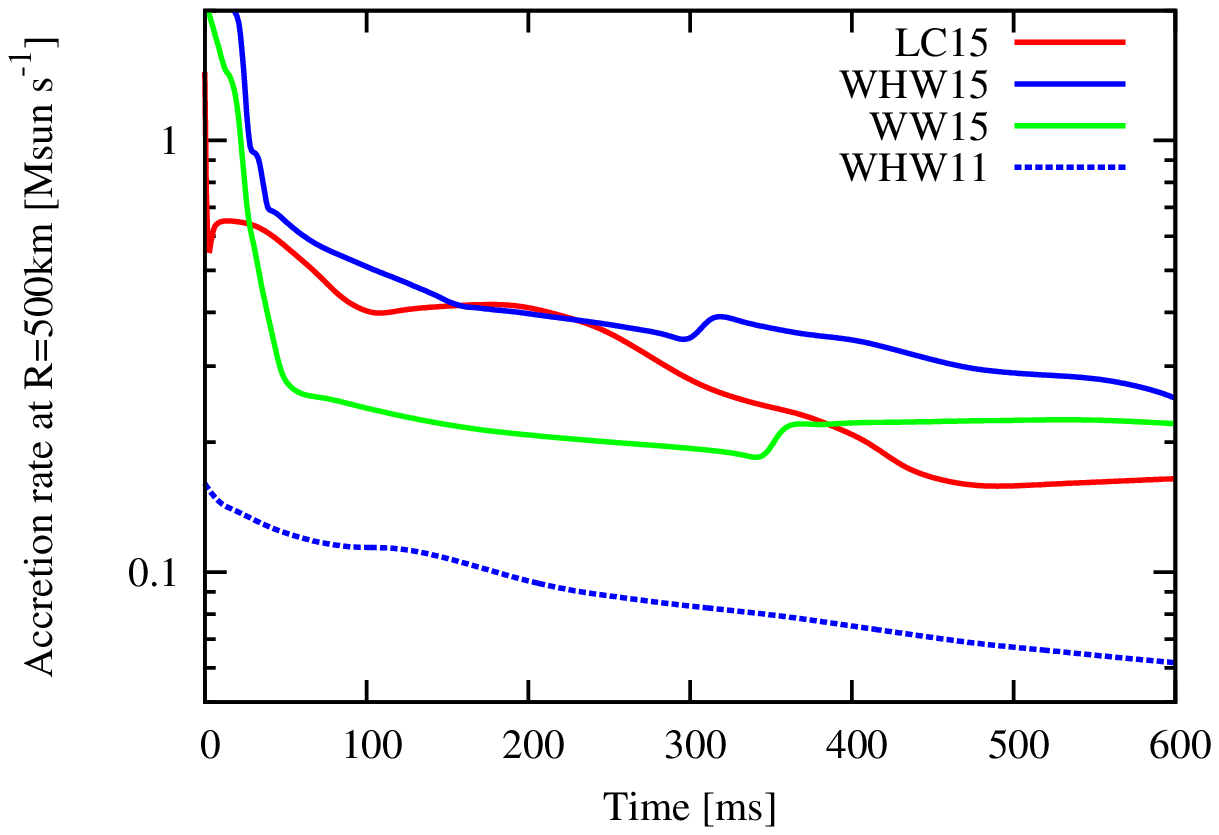}
\end{center}
\caption{
Time evolution of the mass accretion rates, evaluated at $R=500$ km for 4 non-exploding models.
}
\label{fig-macc}
\end{figure}

The critical luminosity for explosions can be read from the $y$-axis in
Figure \ref{fig-para1d} in the limit of long $t_d$ (namely, approaching to a constant neutrino
 luminosity), which corresponds to $2.7$ (WHW15), $2.0$ (LC15),
$1.9$ (WW15), and $1.2$ (WHW11) in unit of $10^{52}$ erg/s,
 respectively. 
The WHW15 model shows the highest critical neutrino luminosity 
among our models. This is because of the higher mass accretion 
rate of the WHW15 model (blue solid line in Figure \ref{fig-macc}), which makes
 the impact of nuclear burning relatively smaller. 
The critical luminosity becomes smallest for model WHW11, 
mainly owing to a compactness of the precollapse core 
and small mass accretion rate coming from its tenuous envelope 
as shown in a blue dotted line in Figure \ref{fig-macc}. 
The mass accretion rate averaged between 200 ms and 600 ms after bounce 
for each model is 
0.33 (WHW15), 0.23 (LC15), 0.21 (WW15), and 0.08 $\Msun~{\rm s}^{-1}$ (WHW11), respectively, 
which is roughly proportional to the critical luminosity except for 
the WHW11 progenitor model. 
When the input luminosity is taken below the critical curves, 
nuclear burning cannot alone drive explosions 
because the shock needs to expand far away from the central protoneutron star 
firstly by neutrino heating
(i.e., the shock revival due to neutrino heating is
 preconditioned to enjoy the assistance from nuclear burning).
 
\begin{figure}[htb]
\begin{center}
\includegraphics[scale=1.2]{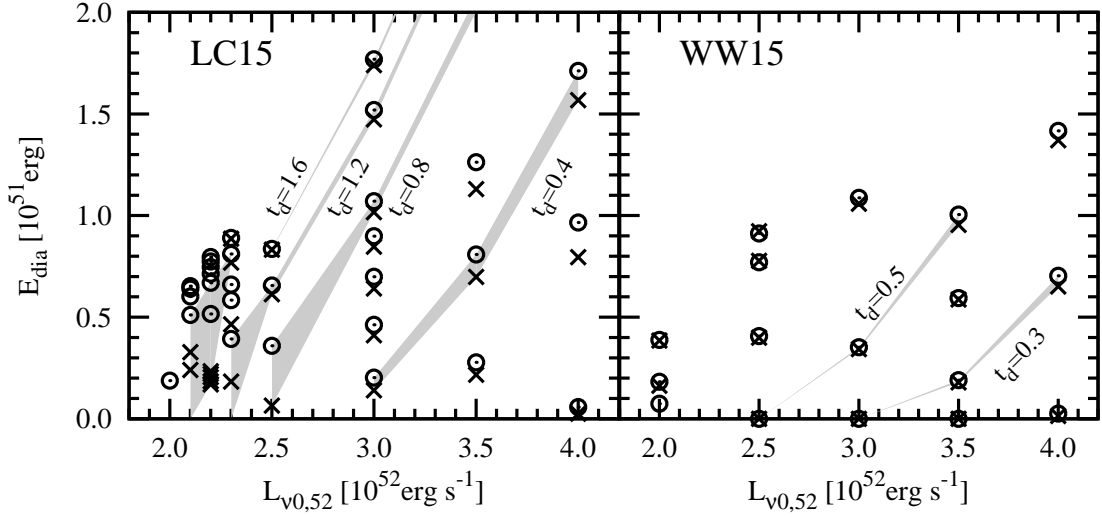}
\caption{
Diagnostic energy as a function of initial neutrino luminosity $L_{\nu0,52}$.
The shaded area shows energy difference between models with ({\it open circles}) and without nuclear burning ({\it crosses}) for some selected $t_{\rm d}$ sequences.
{\it left}: The case of LC 15 progenitor. It can be seen that energy difference 
tends to be large for less energetic models.
{\it right}: The case of WW15 progenitor. The energy difference is small
even for less energetic models.
}
\label{fig-eddif}
\end{center}
\end{figure}

\begin{figure}[htbp]
\begin{center}
\includegraphics[scale=0.8]{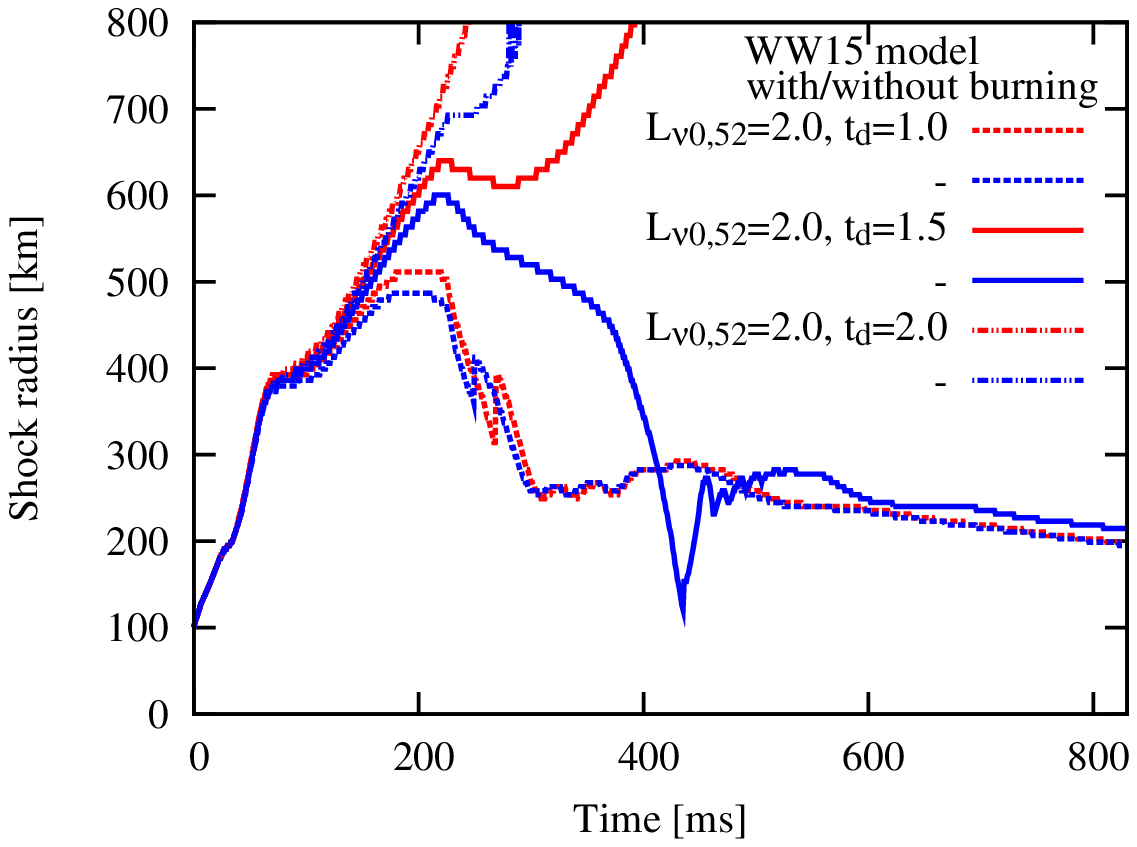}
\end{center}
\caption{Time evolution of the shock radii for several 1D models employing the WW15
progenitor. The shock moves farther out for models including nuclear
burning 
(red lines) compared to those without (blue).}
\label{rsh1d_w15}
\end{figure}

In Table  \ref{tbl-result} time of explosion, $t_{\rm exp}$, and diagnostic energy, $E_{\rm dia}$, are listed for some chosen sets of the neutrino parameters. 
The time of explosion is defined as the moment when a shock reaches an average radius of 4500 km, while non-exploding models are denoted by a ``---'' symbol.
The diagnostic energy is plotted as a function of neutrino luminosity 
for a various decay time scales in Figure \ref{fig-eddif}. 
In fact, the diagnostic energy is shown to be remarkably enhanced in the case of marginal explosions (i.e., low neutrino luminosity $L_{\nu, 0}$ and/or short decay time $t_{d}$), 
and the difference gets small for large $L_{\nu, 0}$ and $t_{d}$, in which explosions are predominantly triggered by neutrino heating. As repeatedly mentioned so far, these features due to nuclear burning are only remarkable in the LC15 progenitor. Nevertheless, it is worth mentioning that 
the shock extent even for the WW95 progenitor, for which
 the impact of nuclear burning is relatively small 
(see the bottom left panel of Figure \ref{fig-para1d}), becomes bigger for models with nuclear burning compared to those without (Figure \ref{rsh1d_w15}). 

\clearpage

\subsection{2D Results}\label{sec-2d}
\begin{figure}[htpb]
\begin{center}
\includegraphics[scale=0.8]{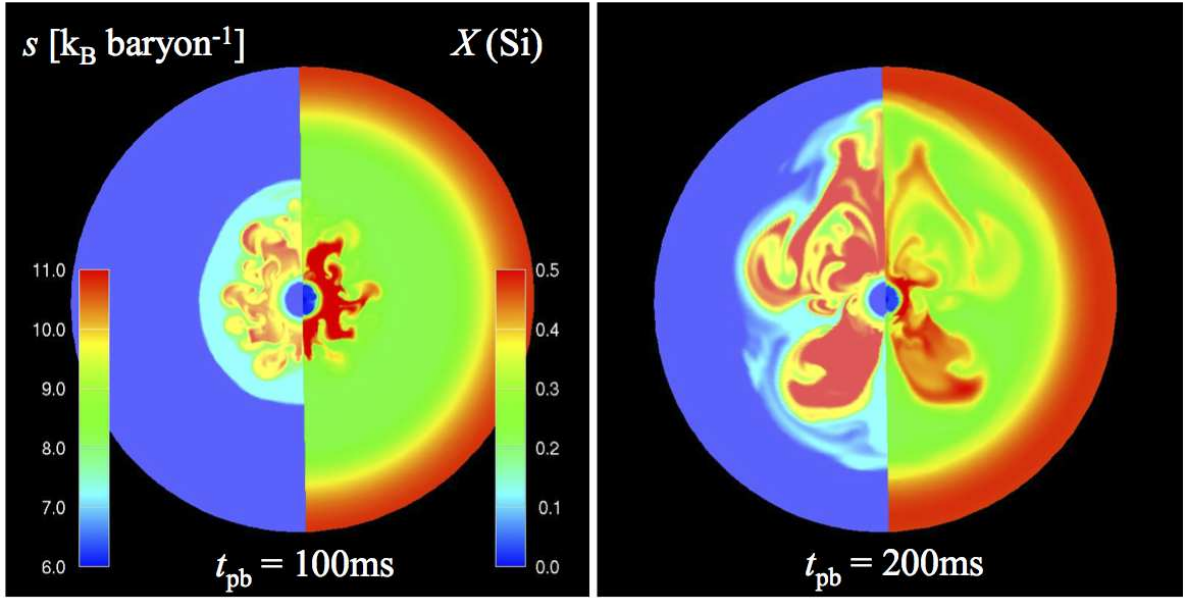}
\end{center}
\caption{
2D distributions of entropy and silicon of LC15 model ($n_{\theta} = 128$). 
Entropy in unit of $k_{\rm B}$ baryon$^{-1}$ is in the left-half sphere 
and silicon mass fraction in the right-half.
Shown is the case of ($L_{\nu0,52}, t_d) = (2.4, 1.1)$ at $t_{\rm pb}$ = 100 ({\it left panel}) and 200 ms ({\it right}) postbounce, respectively.
}
\label{fig-snap2d-1}
\end{figure}

\begin{figure}
\begin{center}
\includegraphics[scale=0.8]{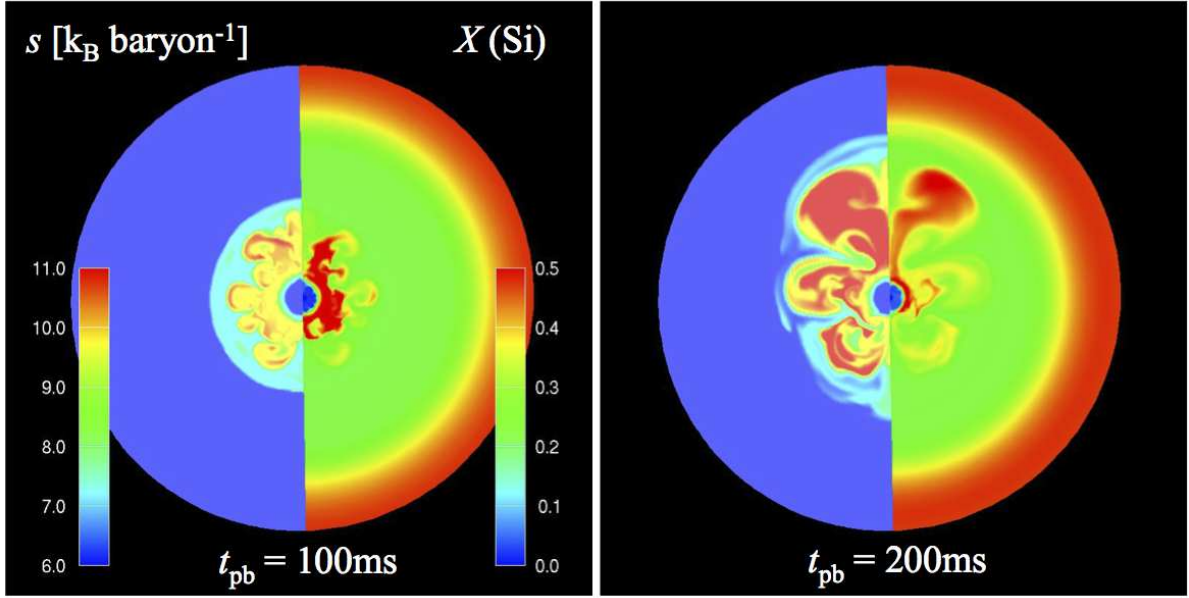}
\end{center}
\caption{Same as Figure \ref{fig-snap2d-1}
but of a 2D model with $(L_{\nu0,52}, t_d)  = (2.2, 1.1)$.}
\label{fig-snap2d-2}
\end{figure}

\begin{figure}[htpb]
\begin{center}
\includegraphics[scale=0.8]{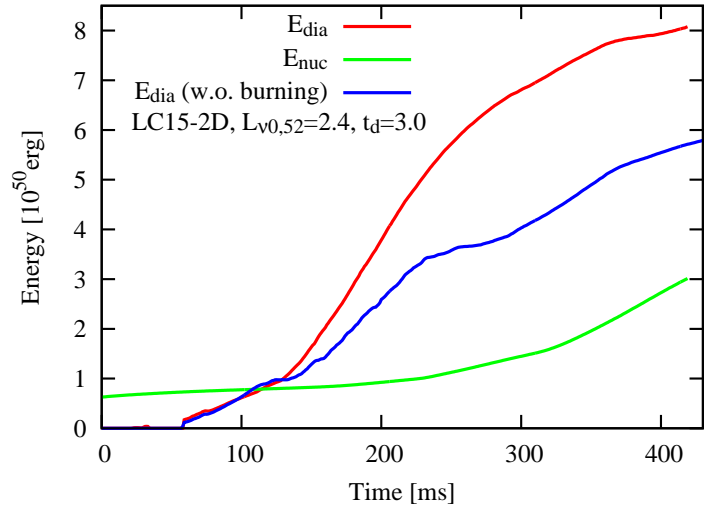}
\end{center}
\caption{Time evolution of the diagnostic energy ({\it red line}) 
and net nuclear burning energy ({\it green}) for LC15 model in 2D ($n_{\theta} = 32$)
with $(L_{\nu 0,52}, t_d) = (2.4, 3.0)$. 
The diagnostic energy of the case without nuclear burning ({\it blue}) is also shown.  
}
\label{fig-eex2d}
\end{figure}

We move on to discuss axi-symmetric 2D models 
and examine the effects of nuclear burning
in the same manner as in the previous section. 
To see clearly the impacts of nuclear burning in our 2D simulations, we
choose to employ the LC15 in the following.

Figures \ref{fig-snap2d-1} and \ref{fig-snap2d-2} show entropy evolution
(left-hand-side in each panel) with the mass fraction of silicon (right-hand-side)
 for two sets of neutrino parameters
at selected postbounce epochs ($t_{\rm pb} = $100 and 200 ms postbounce). 
Small- and large-scale
inhomogeneities in the entropy plots come from neutrino-driven 
convection and
the SASI, both of which lead to more easier explosions in 2D than 1D
\citep[e.g.,][]{Marek09,Murphy08,Ohnishi06}.
 
Reflecting the stochastic motions of the expanding shocks, the way 
how the (anisotropic) shock surfaces touch the nuclear fuel (in the 
shape of spherical shells) changes from model to model in 2D. 
In the case with $L_{\nu0,52}=2.4$ (Figure \ref{fig-snap2d-1}), the expanding 
shock firstly reaches to the silicon layer near in the vicinity of the north pole 
at $t_{\rm} \sim 150$ ms. Simultaneously, heavy elements like nickel 
are synthesized there, which helps to 
push the burning material preferentially along the direction for the moment. 
In a less luminous case assuming smaller luminosity ($L_{\nu0,52}=2.2$), and the same decay time (Figure \ref{fig-snap2d-2}), the shock encounters the silicon layer closer to the center, where a mass accretion rate is effectively higher, resulting in the longer explosion time.

\begin{figure}[htpb]
\begin{center}
\includegraphics{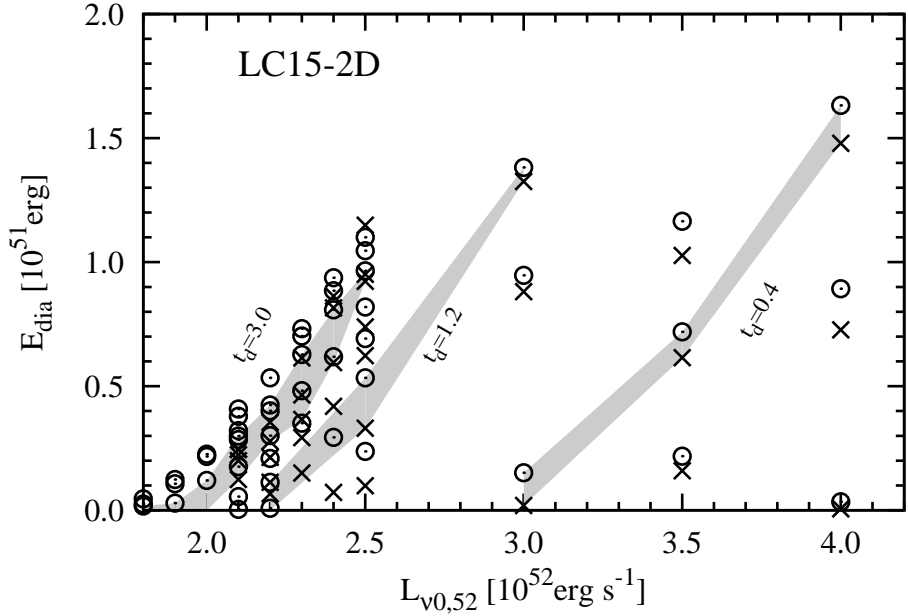}
\end{center}
\caption{
Same as Figure \ref{fig-eddif} but 
for LC15 model in 2D ($n_{\theta} = 32$).
}
\label{fig-eddif2d}
\end{figure}

As shown from Figure \ref{fig-eex2d}, nuclear burning
does assist 2D explosions similar to 1D, but 
 the energy difference (here $\sim 0.2 \times 10^{51}$ erg)
is generally smaller in 2D than in 1D
(compare with Figures \ref{eex1d_m15_2220}, and \ref{eex1d_m15_2250}). 
The comparison of the energy gain due to nuclear burning 
between 1D and 2D models is more clearly shown in 
Table \ref{tbl-result} and Figure \ref{fig-eddif2d}.
The difference of the diagnostic energy with and without nuclear burning 
is larger in 1D than that in 2D.
This may be because neutrino-driven convection and the SASI in 2D models (as 
indicated by entropy distributions in Figures \ref{fig-snap2d-1} and \ref{fig-snap2d-2}) 
enhances 
the neutrino heating efficiency, 
which makes the contribution of nuclear burning relatively 
smaller compare to 1D models.

\begin{figure}[htpb]
\begin{center}
\includegraphics[scale=0.7]{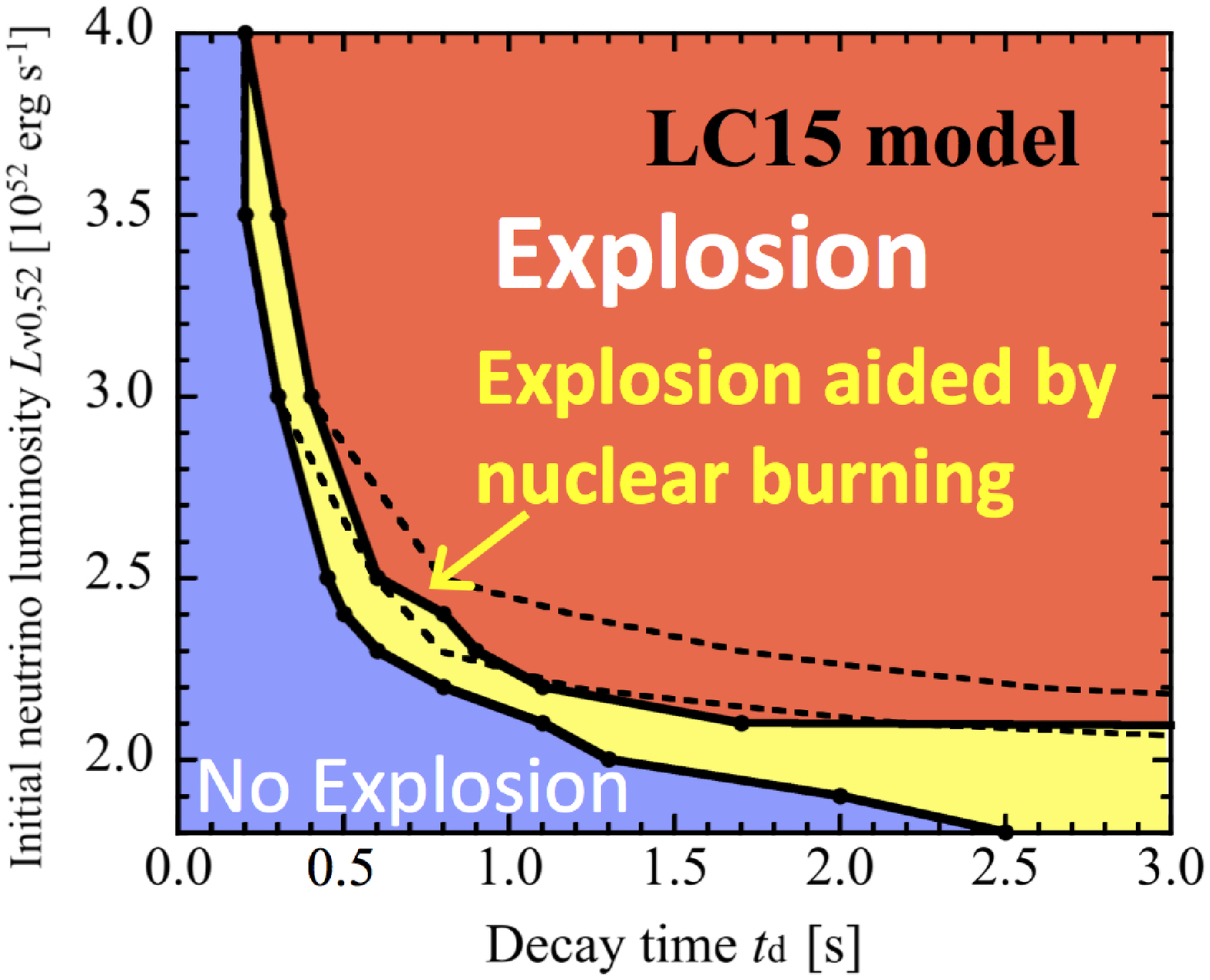}
\end{center}
\caption{Same as Figure \ref{fig-para1d} but for 2D simulations. The
dashed lines represent the critical curves in 1D (compare Figure \ref{fig-para1d}).
 Note that we adopt coarse mesh points in the polar direction (32 uniform grids), so that we can perform 2D simulations for 174 models in total to make this parameter map.
}
\label{fig-para2d}
\end{figure}

Finally Figure \ref{fig-para2d} is the parameter map in 2D 
for the LC15 progenitor.
As expected, 2D hydrodynamics leads to more easier explosions 
compared to 1D (see the dashed lines which are the critical curves in 1D). 
More importantly, the yellow region still exists in 2D models 
for the LC15 progenitor. 
Nuclear energy released in 2D models reduces the critical luminosity
by $0.1-0.5 \times 10^{52} {\rm erg~s^{-1}}$ depending on $t_{\rm d}$,
as well as 1D models, 
although its impact on the diagnostic energy is weaker than for 1D models. 
It would be interesting to perform multi-D (radiation-hydro)
simulations with 
nuclear network calculation for the previously untested progenitor model. 

\section{Conclusions}

We revisited the potential impacts of nuclear burning on 
the onset of neutrino-driven explosions of core-collapse supernovae.
By changing the neutrino luminosity and its decay time
to obtain parametric explosions
in 1D and 2D models with or without a 13-isotope $\alpha$ network, 
we studied how
the inclusion of nuclear burning could affect
 the postbounce dynamics for four progenitor models; three for $15.0
 \Msun$ stars of 
 \citet{limongi}, \citet{WW95}, and \citet{woos02},
 and one for an $11.2 \Msun$ star of \citet{woos02}
 Our results showed that the energy gain due to nuclear burning of 
infalling material behind the shock can energize the
shock expansion especially for models that produce only marginal explosions
 in the absence of nuclear burning. These models enjoy
the assistance from nuclear burning typically in the
following two ways, whether the shock front passes through the
 silicon-rich layer, or later it touches to the oxygen-rich layer.
 Depending on the neutrino luminosity and its decay time,
 the diagnostic energy of explosion was found to increase up to a few times $10^{50}$ erg 
 for models with nuclear burning compared to the corresponding models without.
 The energy difference becomes generally smaller in 2D
than in 1D, because 
neutrino-driven convection and the SASI in 2D models enhance the neutrino heating efficiency, making the contribution of nuclear burning relatively smaller compared to 1D models. 
It was pointed out that these features are most remarkable for the LC15 progenitor, 
which possesses a massive oxygen layer
with its inner-edge radius
being smallest among the employed progenitors, 
which makes the timescale shorter for the shock to encounter the rich fuel.
Considering reduction of the critical luminosity 
and increase of the diagnostic energy by nuclear burning, 
and also uncertainties in the structure of progenitors, 
our results indicate
that nuclear burning should still remain as one of the important
ingredients to foster the onset of neutrino-driven explosions.

\acknowledgments
We thank H.T. Janka for stimulating discussions and we are also 
grateful to T. Kuroda, Y. Suwa, T. Kajino for helpful exchanges.
KK and TT are thankful to S. Yamada and K. Sato for continuing encouragements.
NN was financially supported by the European Research Council under
EU-FP7-ERC-2012-St Grant 306901.
Numerical computations were carried out in part on XT4 and 
general common use computer system at the Center for Computational Astrophysics, CfCA, 
the National Astronomical Observatory of Japan.  This 
study was supported in part by the Grants-in-Aid for the Scientific Research 
from the Ministry of Education, Science and Culture of Japan
(Nos. 20740150, 23340069, and 23540323)
and by HPCI Strategic Program of Japanese MEXT.

\begin{table}[htbp]
\caption{Summary of results.}
\begin{tabular}{ccccccccc}
\hline \hline
                         &&&  & \multicolumn{2}{c}{without burning} && \multicolumn{2}{c}{with burning}\\
                          \cline{5-6}  \cline{8-9}
model & $L_{\nu 0,52}$ & $t_{\rm d}$ && $t_{\rm exp}$ & $E_{\rm dia}$ && $t_{\rm exp}$& $E_{\rm dia}$\\
       & ($10^{52}{\rm erg \, s}^{-1}$) & (s)   &&       (s)           & ($10^{51}$erg)  &&   (s)       & ($10^{51}$erg)\\
\hline
LC15             & 2.0 &    5.0           && ---                      & ---                         && 0.836      & 0.188                      \\
                     & 2.2 &    2.0            && ---                      & ---                         && 0.502      & 0.516            \\
                     & 2.2 &    5.0            && 0.819                & 0.234                     && 0.465      & 0.796             \\
                     & 2.5 &    0.8            && 0.811                & 0.065                     && 0.460      & 0.359              \\
                     & 2.5 &    1.6            && 0.462                & 0.832                     && 0.416      & 0.836                \\
                     & 3.0 &    0.4            && 0.498                & 0.140                     && 0.424      & 0.203                \\
                     & 3.0 &    0.8            && 0.394                & 1.017                     && 0.368      & 1.071                \\
WHW15      & 2.7 &    2.5            && 0.712                & 0.765                      && 0.598      & 0.906                      \\
                     & 3.0 &    1.0            && 0.556                & 0.613                    && 0.531      & 0.644                      \\
                     & 3.5 &    0.5            && 0.493                & 0.379                     && 0.477      & 0.403                      \\
                     & 4.0 &    0.4            && 0.455                & 0.622                    && 0.403      & 0.706                      \\
WW15         & 2.0 &    5.0            && 0.543                & 0.385                     && 0.527      & 0.388                      \\
                     & 3.0 &    1.0            && 0.360                & 1.058                     && 0.357      & 1.087                      \\
                     & 3.5 &    0.5            && 0.336                & 0.954                     && 0.333      & 1.005                      \\
                     & 4.0 &    0.3            && 0.325                & 0.651                     && 0.322      & 0.704                      \\
WHW11      & 1.5 &    5.0            && 0.472                & 0.195                      && 0.465      & 0.208                      \\
                     & 2.0 &    5.0            && 0.353                & 0.821                     && 0.349      & 0.845                      \\
                     & 2.5 &    0.7            && 0.319                & 0.800                     && 0.316      & 0.837                      \\
                     & 3.0 &    0.4            && 0.292                & 0.911                     && 0.289      & 0.951                      \\
\hline
LC15 (2D)  & 2.0 &    4.0            && ---                      & ---                            && 0.541      & 0.226                      \\
                     & 2.2 &    2.0            && 0.664                & 0.113                     && 0.491      & 0.301           \\
                     & 2.2 &    4.0            && 0.542                & 0.355                     && 0.488      & 0.534            \\
                     & 2.5 &    0.8            && 0.596                & 0.099                     && 0.461      & 0.237             \\
                     & 2.5 &    1.6            && 0.464                & 0.623                     && 0.425      & 0.692               \\
                     & 3.0 &    0.4            && 0.514                & 0.019                     && 0.437      & 0.151               \\
                     & 3.0 &    0.8            && 0.401                & 0.881                     && 0.372      & 0.947               \\
\hline \hline 
\\
\\
\end{tabular}
\label{tbl-result}
\end{table}


\appendix

\section{$\alpha$ network versus flashing method}\label{app-flash}
\begin{figure}[htbp]
\includegraphics{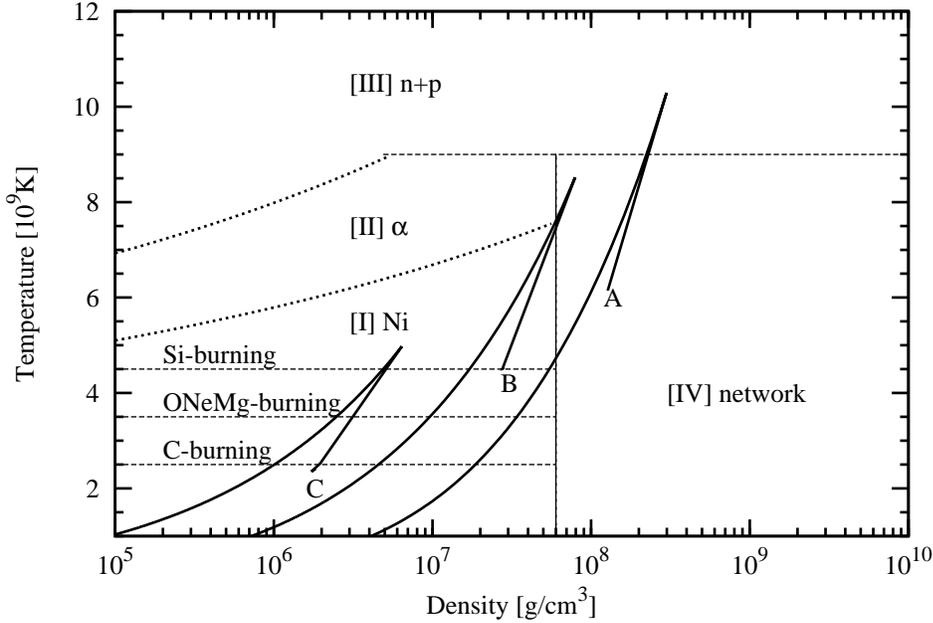}
\caption{
A Schematic picture of $\rho - T$ plane characterizing our pseud-flusing method. 
Region I almost consists of nickel, region II contains $\alpha$-particles and free nucleons, and in region III all nuclei and $\alpha$-particles are dissolved into free nucleons.
In region IV we resolve the nuclear network to evolve the chemical compositions.
See \citet{rampp02} for details.
We take 3 flows (A, B, and C) as representations of mass shells in LC 15 model.
}
\label{fig-dpplane}
\end{figure}

In this paper, we use a reaction network involving 13 $\alpha$ nuclei for the purpose to investigate a potential role of nuclear burning in reviving and strengthening weak shocks in neutrino-driven 
explosions.
Although the 13-$\alpha$ network calculation itself is rather 
simple, it is computationally expensive to perform them for 
each species evolved with multi-D hydrodynamics.
To avoid this expense, various simplifications are employed, for example, a ``flashing method'' \citep{rampp02}.
In the flashing method, a hydrodynamic flow is characterized by its matter density $\rho$ and temperature $T$.
The flow travels in a $\rho$-$T$ plane and changes its chemical composition and releases nuclear energy according to the region where it is in the plane.
Figure \ref{fig-dpplane} presents the composition of flows in the $\rho$-$T$ plane, which is slightly different from the original one \citep{rampp02}. 
We compare the evolution of composition and subsequent energy release of flows in this ``pseudo-" flashing method with those in our $\alpha$  network.

Following to  \citet{rampp02} we assume that 
dissociation of nuclei and the recombination of free nucleons and $\alpha$-particles
change the chemical composition 
below the transition density ($\rho_0 = 6 \times 10^7 {\rm g \, cm^{-3}}$).
In region I all free nucleons and $\alpha$ particles form $^{56}$Ni.
In region II all heavy nuclei are dissolved and free nucleons recombine into $\alpha$ particles.
In region III all heavy nuclei and $\alpha$ particles are disintegrated into free nucleons.
These three regions are separated by two curves $\rho_1(T), \, \rho_2(T)$ in the $\rho - T$ plane:
\begin{equation}
\log_{10}(\rho_1(T)) = 11.62 + 1.5 \log_{10}(T_9) - 39.17/T_9,
\end{equation}
\begin{equation}
\log_{10}(\rho_2(T)) = 10.60 + 1.5 \log_{10}(T_9) - 46.54/T_9,
\end{equation}
where $T_9$ is the temperature in unit of $10^9$ K.
Above the transition density (region IV) we calculate the nuclear network 
instead of the use of the equation of state of \cite{latt91} as in \citet{rampp02}.
Three horizontal lines at $T_9=2.5, \, 3.5,$ and $4.5$ present 
$^{12}$C burning to $^{24}$Mg, $^{16}$O-$^{20}$Ne-$^{24}$Mg burning to $^{28}$Si,
and $^{28}$Si burning to $^{56}$Ni, respectively.
At $T_9 > 9$ we assume that all nuclei are disintegrated into free nucleon independent of density.

\begin{figure}[htbp]
\includegraphics[scale=1.2]{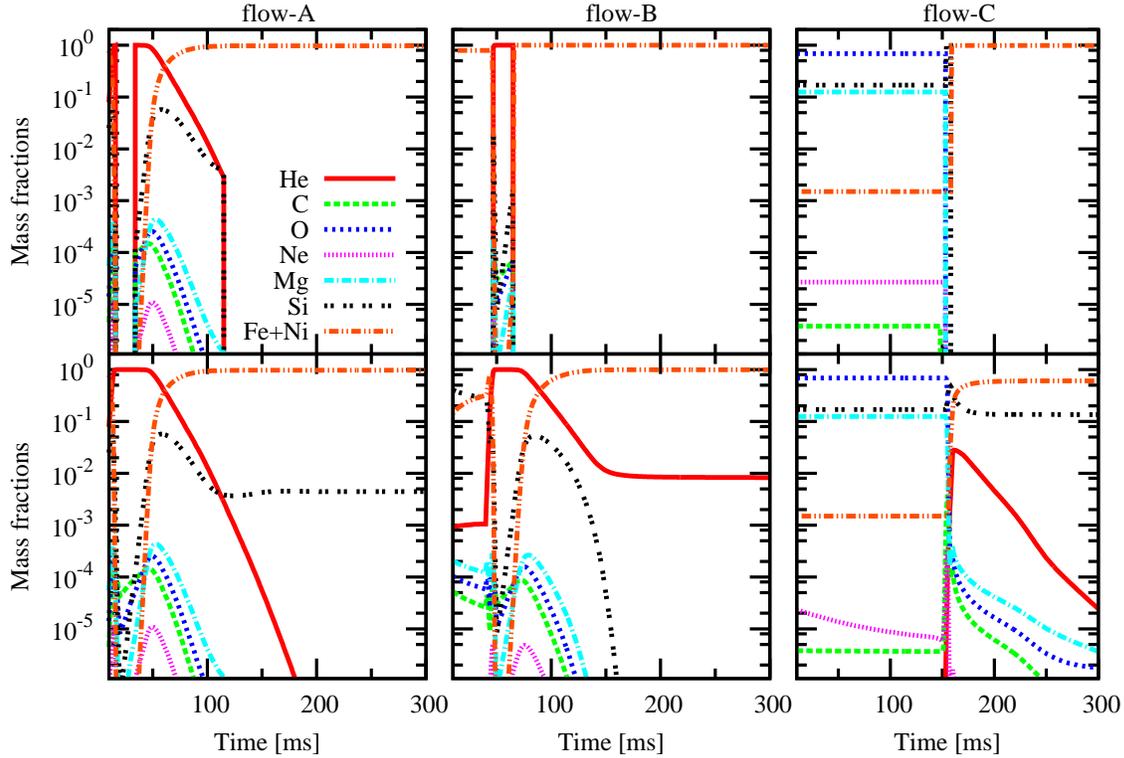}
\caption{Time evolution of chemical compositions of flow-A, B and C.
Shown are the results from pseudo-flashing method ({\it top panels}) and from network calculation ({\it bottom}).
}
\label{fig-abund}
\end{figure}

\begin{figure}[htbp]
\includegraphics{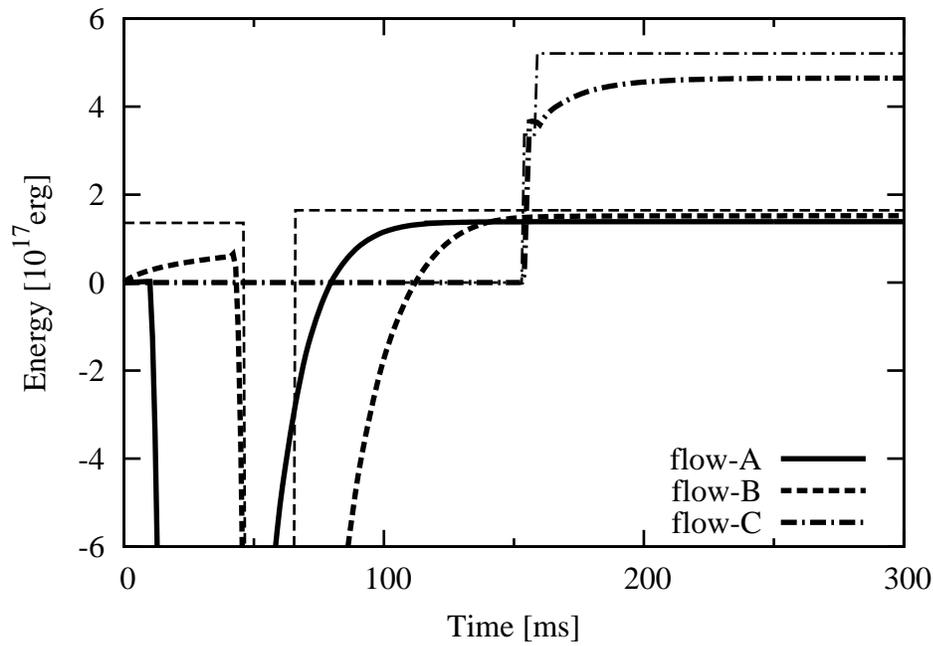}
\caption{
Temporal summations of released energy through nuclear reactions.
Results form network calculation ({\it thick lines}) and pseudo-flashing method ({\it thin lines}) are shown for 3 flows.
The solid thin line (flow-A calculated with pseudo-flashing method) presents almost the same evolution as the network case and is hidden by the solid thick line.
}
\label{fig-energy}
\end{figure}

Here we take three flows in LC 15 model named A, B, and C, located in the mass coordinate at 
$1.3 \Msun$ (Fe core), $1.5 \Msun$ (Si layer), and $1.7 \Msun$ (O/Si layer), respectively.
These flows are launched via 1-dimensional hydrodynamic simulation by putting thermal energy 
in the innermost region of the iron core so that the explosion energy of $10^{51}$ erg is obtained.
Each flow undergoes shock heating and compression (to the upper right direction in Figure \ref{fig-dpplane}),
then expands and gets cool (to the lower left).
We compare the change of chemical compositions (Figure \ref{fig-abund}) and released energy  (Figure \ref{fig-energy}) via the flashing method with those estimated by $\alpha$ network.
Both of the methods show similar energy yield for all flows at late phase, although the intermediate evolution and final abundance of intermediate-mass elements are different.
This difference is caused by our treatment that the composition immediately changes when a flow goes across a line separating regions presented in Figure \ref{fig-dpplane}.
To avoid the unrealistic jump, \citet{rampp02} introduced factors $f_{\rm I}, \, f_{\rm II}$, and $f_{\rm III}$
so that the composition change progresses gradually.
We conclude that the flashing method is a good approximation and useful for SN simulations.

\section{Light-bulb scheme versus isotropic diffusion source approximation}
Our current study is based on the light-bulb (LB) scheme,
 in which a prescribed neutrino heating and cooling rate is 
 used. This simple approach reduces computational cost 
compared with more sophisticated simulations
and makes our extensive parameter study possible.
The light-bulb approximation has been frequently used for various 
purposes, such as to study effects of spacial 
dimensionality \citep[e.g.,][]{dolence13,couch12} and progenitor 
inhomogeneities \citep{couch13b} on the neutrino-driven mechanism, explosive 
nucleosynthesis \citep[e.g.,][]{fujimoto}, 
gravitational-wave signals \citep[e.g.,][]{kotake09a,kotake09b,kotake11a}, and so on.
The light-bulb method is also useful for removing feedbacks from 
physical inputs into the neutrino luminosity and temperature,
which enables to investigate the relative changes 
due to different choice of the physical inputs 
\citep[e.g., EOS study by][]{couch13}.

\begin{figure}[htbp]
\plottwo{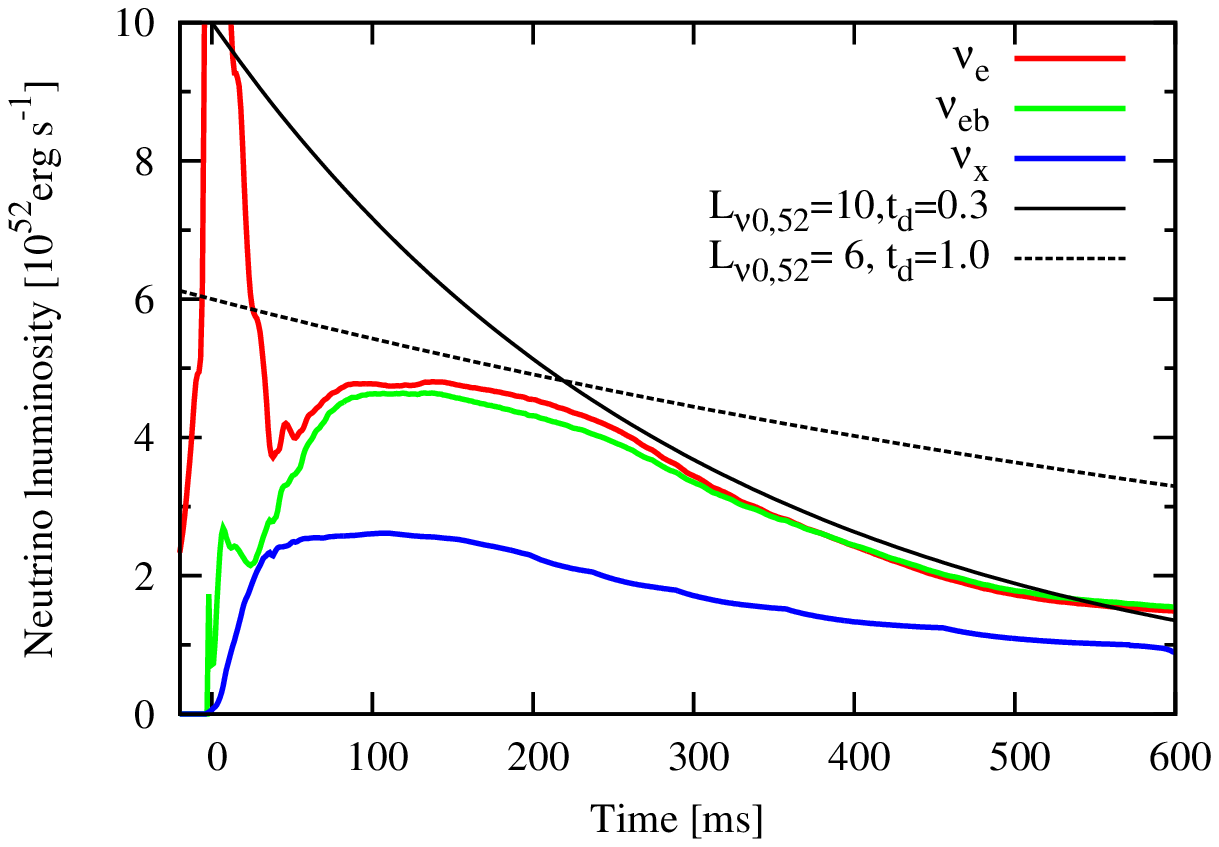}{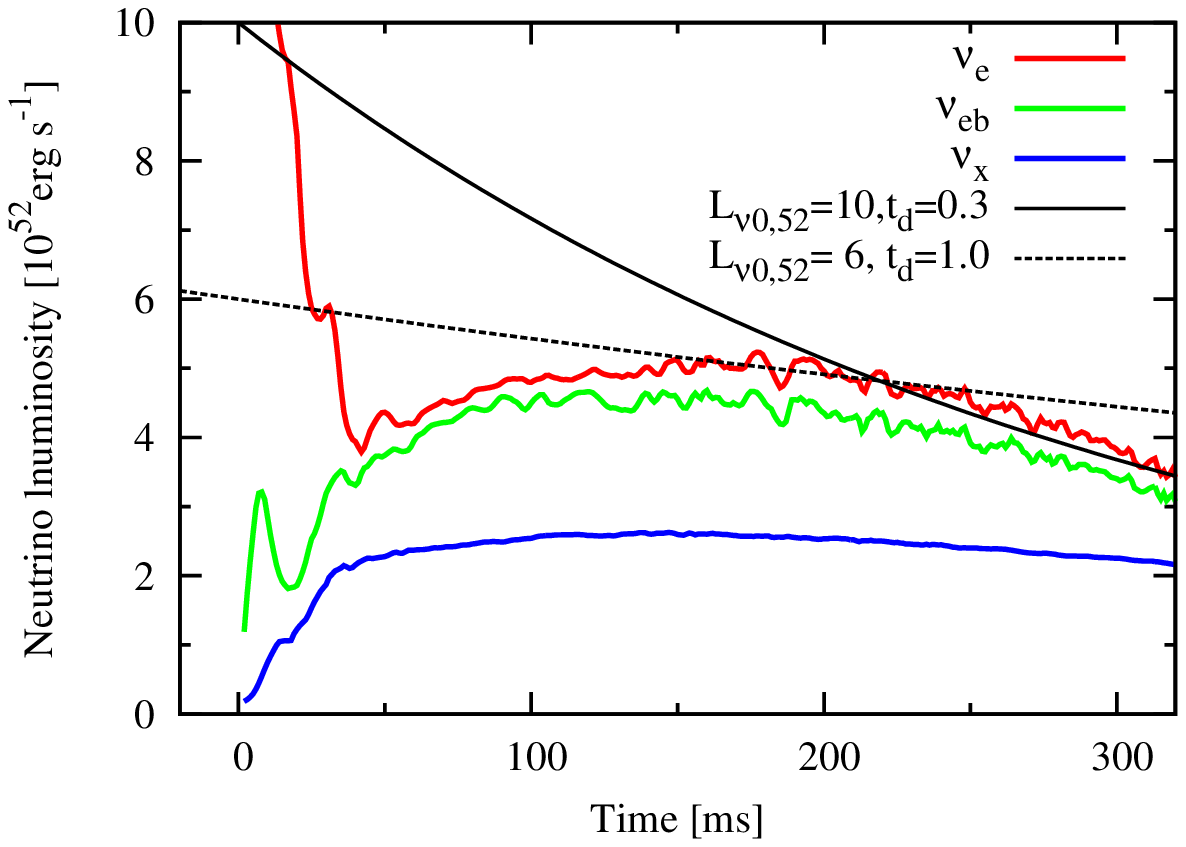}
\caption{
Neutrino luminosity resulting from IDSA simulations of 1D (left) and 2D (right) for LC15 model.
}
\label{idsa-ln}
\end{figure}

\begin{figure}[htbp]
\begin{center}
\includegraphics{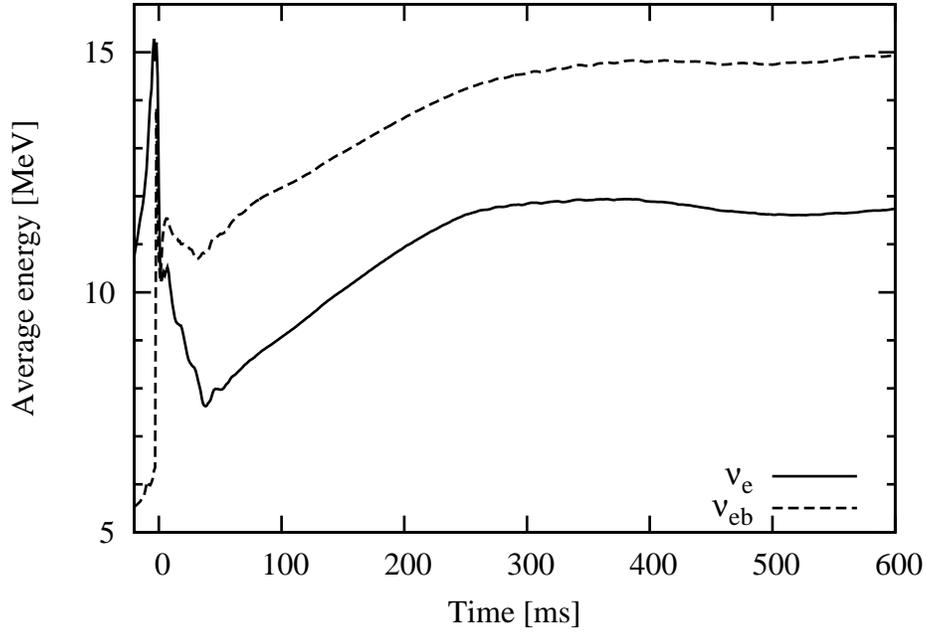}
\end{center}
\caption{
Time evolution of neutrino energy for 1D model.
}
\label{idsa-en}
\end{figure}

In this appendix, we briefly discuss the validity of the LB scheme by
  comparing the neutrino luminosities, average energy, 
and heating rates assumed in this study with those 
 from 1D simulations in which spectral neutrino transport is solved 
 by the isotropic diffusion source approximation (IDSA) scheme (see,
 \citet{idsa} for more detail). We employ the LC15 progenitor model 
 in both of the two runs.
Figure \ref{idsa-ln} shows the time evolution of neutrino luminosities estimated from IDSA simulations.
In the previous studies using the light-bulb models, the neutrino luminosities are assumed to be constant, although it is apparently unrealistic.
In this paper, we introduced another parameter, the decay time of neutrino luminosities, to complement this discrepancy.
The neutrino luminosities are not a monotonic function of time and only a late phase can be fit by an exponential decay 
with $t_{\rm d}=0.3$ s.
The evolution of average energy of neutrinos is shown in Figure \ref{idsa-en}.
According to the previous studies, we assume the electron neutrino temperature to be kept constant as $4$ MeV.
Figure  \ref{idsa-en} presents an almost constant energy of neutrinos with 
$E_{\nu_{\rm e}} \sim 12$MeV, indicating that the constant neutrino temperature is not a 
 bad assumption.
Finally we compare the neutrino heating rate from the 
IDSA simulations with that from the light-bulb scheme (Figure 24). Taking 
the neutrino luminosity, electron and neutron fractions, and temperature 
distributions from the IDSA simulation, we put them into
Eqs. (\ref{eq-heating}) and (\ref{eq-cooling}) to estimate the heating 
 rate by the light-bulb scheme. Here the neutrino temperature is 
fixed to be $4$ MeV. 
Note that we drop an suppression term $e^{\tau_{\nu_{\rm e}}}$ for the current estimation.
The light-bulb model captures well the heating and cooling regions, 
although the heating rate behind the shock is underestimated compared to that from IDSA.
Nevertheless, it is still a powerful tool for a parametric search to explore qualitative trends,
such as effects of neutrino luminosity, its decay timescale, and nuclear burning on boosting the onset of neutrino-driven explosions.

\begin{figure}[htbp]
\includegraphics[scale=1.2]{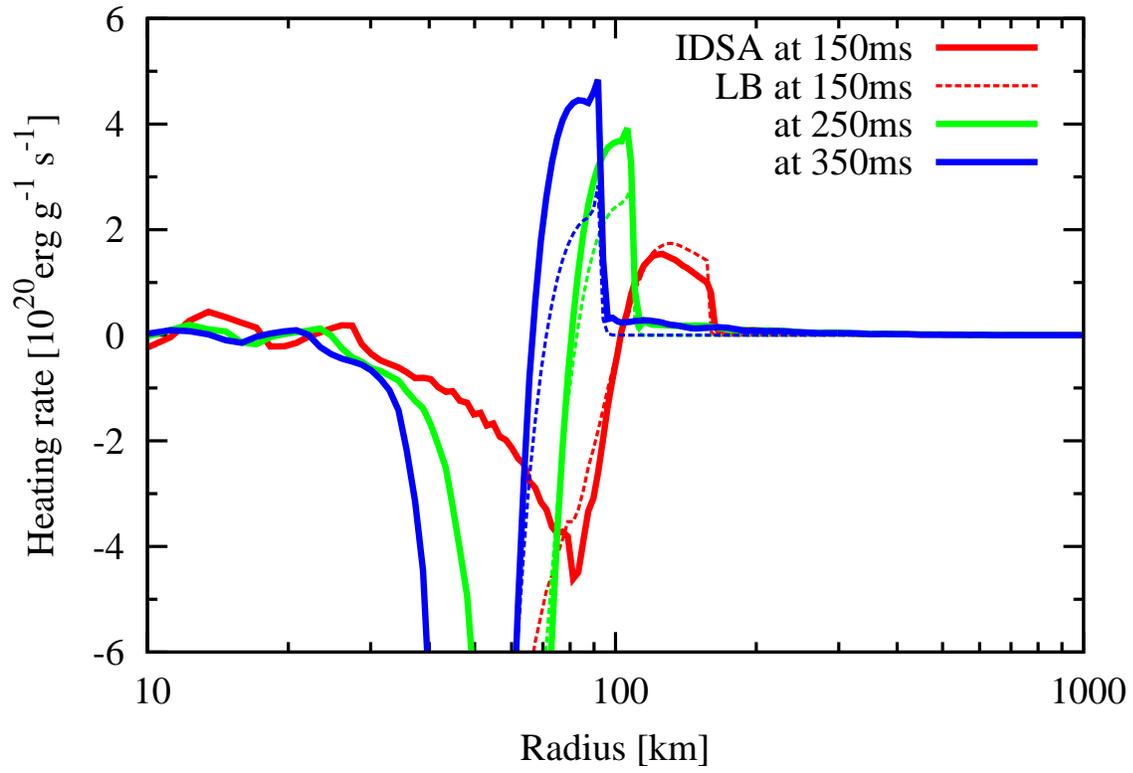}
\caption{
Radial distribution of heating rates for 1D model at $t_{\rm pb}=$150 (red lines), 250 (green), and 350 ms (blue). LB models (thin lines) tend to overestimate the heating rate in a pre-shock region and underestimate in a post-shock region.
}
\label{idsa-heat}
\end{figure}

\end{document}